\documentclass[]{article}
\usepackage{lmodern}
\usepackage{amssymb,amsmath}
\usepackage{ifxetex,ifluatex}
\usepackage{fixltx2e} 
\ifnum 0\ifxetex 1\fi\ifluatex 1\fi=0 
  \usepackage[T1]{fontenc}
  \usepackage[utf8]{inputenc}
\else 
  \ifxetex
    \usepackage{mathspec}
  \else
    \usepackage{fontspec}
  \fi
  \defaultfontfeatures{Ligatures=TeX,Scale=MatchLowercase}
\fi
\IfFileExists{upquote.sty}{\usepackage{upquote}}{}
\IfFileExists{microtype.sty}{%
\usepackage{microtype}
\UseMicrotypeSet[protrusion]{basicmath} 
}{}
\usepackage[margin=4cm]{geometry}
\usepackage[hidelinks]{hyperref}
\hypersetup{unicode=true,
            pdftitle={Structural Equation Models as Computation Graphs},
            pdfauthor={Erik-Jan van Kesteren and Daniel Oberski},
            pdfborder={0 0 0},
            breaklinks=true}
\usepackage{longtable,booktabs}
\usepackage{graphicx,grffile}
\makeatletter
\def\maxwidth{\ifdim\Gin@nat@width>\linewidth\linewidth\else\Gin@nat@width\fi}
\def\maxheight{\ifdim\Gin@nat@height>\textheight\textheight\else\Gin@nat@height\fi}
\makeatother
\setkeys{Gin}{width=\maxwidth,height=\maxheight,keepaspectratio}
\IfFileExists{parskip.sty}{%
\usepackage{parskip}
}{
\setlength{\parindent}{0pt}
\setlength{\parskip}{6pt plus 2pt minus 1pt}
}
\ifx\paragraph\undefined\else
\let\oldparagraph\paragraph
\renewcommand{\paragraph}[1]{\oldparagraph{#1}\mbox{}}
\fi
\ifx\subparagraph\undefined\else
\let\oldsubparagraph\subparagraph
\renewcommand{\subparagraph}[1]{\oldsubparagraph{#1}\mbox{}}
\fi

\let\rmarkdownfootnote\footnote%
\def\footnote{\protect\rmarkdownfootnote}

\usepackage{titling}

  \title{Structural Equation Models as Computation Graphs}
    \pretitle{\vspace{\droptitle}\centering\huge}
  \posttitle{\par}
    \author{Erik-Jan van Kesteren and Daniel Oberski}
    \preauthor{\centering\large\emph}
  \postauthor{\par}
      \predate{\centering\large\emph}
  \postdate{\par}
    \date{Utrecht University, Department of Methodology and Statistics}

\usepackage{booktabs}
\usepackage{longtable}
\usepackage{array}
\usepackage{multirow}
\usepackage{wrapfig}
\usepackage{float}
\usepackage{colortbl}
\usepackage{pdflscape}
\usepackage{tabu}
\usepackage{threeparttable}
\usepackage{threeparttablex}
\usepackage[normalem]{ulem}
\usepackage{makecell}
\usepackage{xcolor}

\usepackage{float}
\usepackage{caption}
\begin{document}
\maketitle
\begin{abstract}
Structural equation modeling (SEM) is a popular tool in the social and behavioural sciences, where it is being applied to ever more complex data types. The high-dimensional data produced by modern sensors, brain images, or (epi)genetic measurements require variable selection using parameter penalization; experimental models combining disparate data sources benefit from regularization to obtain a stable result; and genomic SEM or network models lead to alternative objective functions. With each proposed extension, researchers currently have to completely reformulate SEM and its optimization algorithm -- a challenging and time-consuming task.

In this paper, we consider each SEM as a \emph{computation graph}, a flexible method of specifying objective functions borrowed from the field of deep learning. When combined with state-of-the-art optimizers, our computation graph approach can extend SEM without the need for bespoke software development. We show that both existing and novel SEM improvements follow naturally from our approach. To demonstrate, we discuss least absolute deviation estimation and penalized regression models. We also introduce spike-and-slab SEM, which may perform better when shrinkage of large factor loadings is not desired. By applying computation graphs to SEM, we hope to greatly accelerate the process of developing SEM techniques, paving the way for new applications. We provide an accompanying \texttt{R} package \texttt{tensorsem}.
\end{abstract}

{
\setcounter{tocdepth}{2}
\tableofcontents
}
\hypertarget{introduction}{%
\section{Introduction}\label{introduction}}

Structural equation modeling (SEM) is a popular tool in the social and behavioural sciences, where it is being applied to ever more complex data types. SEM extensions now perform variable selection in high-dimensional situations (Jacobucci, Brandmaier, \& Kievit, 2018; van Kesteren \& Oberski, 2019), modeling of intensive longitudinal data (Asparouhov, Hamaker, \& Muthén, 2018; Voelkle \& Oud, 2013), and analysis of intricate online survey experiments (Cernat \& Oberski, 2019). In these situations, the SEM model often needs to be reformulated and traditional optimization approaches need to be extended to obtain parameter estimates -- a challenging and time-consuming task. For example, applying SEM to high-dimensional data warrants parameter penalization, and special model types such as genomic SEM (Grotzinger et al., 2019) or network models (Epskamp, Rhemtulla, \& Borsboom, 2017) can lead to alternative objective functions. Additionally, even before the extension of SEM to novel data structures there have been several examples of the instability of the latent variable approach -- such as Heywood cases (Kolenikov \& Bollen, 2012) and convergence problems in multitrait-multimethod (MTMM) models (Revilla \& Saris, 2013), which may benefit from regularization to obtain a stable result. As SEM expands to ever larger, more complex applications, the estimation challenge grows and currently available methods will be insufficient.

To support the evolution of SEM, we see an opportunity to use existing solutions from the field of deep learning, which has been pioneering methods for overparameterized, unstable, complex models for decades (see Goodfellow, Bengio, \& Courville, 2016 and the references therein for an overview). In order to make use of the progress from this field, there is a need for a formulation of SEM with latent variables that allows researchers to implement these existing solutions directly. To this end, this paper introduces computation graphs as a method to specify and extend SEM, enabling flexible editing of the objective function used to estimate its parameters. We show that improvements possible through the computation graph approach include -- but are not limited to -- extensions to SEM that have been proposed in the literature, such as least absolute deviation estimation (Siemsen \& Bollen, 2007), LASSO regularized regression (Friedman, Hastie, \& Tibshirani, 2010; Jacobucci, Grimm, \& McArdle, 2016), and sparse loading estimation in factor analysis (Jin, Moustaki, \& Yang-Wallentin, 2018). For this last extension, we introduce the spike-and-slab penalty (Ročková \& George, 2018) in SEM. Relative to L1 and L2 penalties, this shrinks large coefficients less, inducing less bias in the most interesting parameters. Our approach makes it relatively easy to import such solutions from literature in other fields.

In the following, SEM will first be framed as an optimization problem, and a brief overview will be given of the current methods of SEM parameter estimation and their limitations. Then, we will introduce the concept of computation graphs, as used in the field of deep learning, along with an introduction to the optimizers intended to deal with these limitations. Subsequently, we will develop the computation graph for SEM, after which we show both theoretically and practically how this can be used to extend SEM to novel situations. In doing so, we also introduce the spike-and-slab penalization as a promising method for SEM regularization, and as an example of the ease with which SEM model can be extended using our framework. Lastly, we discuss the implications of this novel framework for SEM and we provide directions for future research. The methods introduced this paper are provided as an open-source \texttt{R} (R Core Team, 2018) package called \texttt{tensorsem} (van Kesteren, 2019), which is available at \href{https://github.com/vankesteren/tensorsem}{github.com/vankesteren/tensorsem}. All the examples and simulations associated with this paper are reproducible using the stable \texttt{computationgraph} git branch of the software project.

\hypertarget{background}{%
\section{Background}\label{background}}

\hypertarget{sem-as-an-optimization-problem}{%
\subsection{SEM as an optimization problem}\label{sem-as-an-optimization-problem}}

SEM in its basic form (Bollen, 1989) is a framework to model the covariance matrix of a set of observed variables. Through separation of structural and measurement models, it enables a wide range of multivariate models with both observed and latent variables. SEM generalizes many common data analysis methods, such as linear regression, seemingly unrelated regression, errors-in-variables models, confirmatory and exploratory factor analysis (CFA / EFA), multiple indicators multiple causes (MIMIC) models, instrumental variable models, random effects models, and more.

A key contribution of SEM is the development of a language to specify matrices that hold the parameters \(\boldsymbol{\theta}\), and methods to compute their estimates. The development of these matrix formulations has allowed users to specify models flexibly, intuitively, and even graphically in \texttt{lavaan} (Rosseel, 2012), Onyx (Oertzen, Brandmaier, \& Tsang, 2015), LISREL (Jöreskog \& Sörbom, 1993), Stata (StataCorp, 2017), MPlus (Muthén \& Muthén, 2019), and IBM SPSS AMOS (Byrne, 2016). This availability of user-friendly software has undoubtedly played an important role in the current widespread adoption of SEM in practice. Models may include any combination of latent variables, regression paths, and factor loadings, but even in complex models practioners are shielded from the computational problem of arriving at parameter estimates based on their data -- assuming the model has been well-specified and the data allow for its identification.

Below, we explain how the SEM matrix notation leads to a model-implied covariance matrix. Then, we show how this matrix is the basis for objective functions representing the distance between the model-implied and the observed covariance matrix. Next, we show how this objective is used to estimate of the parameters of interest in the maximum likelihood (ML) and generalized least squares (GLS) frameworks. Last, we discuss the limits of this framework in the context of increasing model complexity.

\hypertarget{notation}{%
\subsubsection{Notation}\label{notation}}

The most commonly used languages are the LISREL notation (Jöreskog \& Sörbom, 1993) used in software packages such as \texttt{lavaan} (Rosseel, 2012) and the Reticular Action Model (RAM) notation (McArdle \& McDonald, 1984) used in software such as OpenMX (Neale et al., 2016). In this paper, we adopt a variant of the LISREL notation used in \texttt{lavaan}, also known as the ``all-y'' version. We consider the measurement and structural equations as follows:
\begin{alignat}{2}
\boldsymbol{z} &= \boldsymbol{\Lambda\eta} + \boldsymbol{\varepsilon} &&\text{(Measurement model)}\\ 
\boldsymbol{\eta} &= \boldsymbol{B}_0\boldsymbol{\eta} + \boldsymbol{\xi} \quad &&\text{(Structural model)}
\label{eq:semeq}
\end{alignat}

where \(\boldsymbol{z}\) represents a vector of centered observable variables of length \(P\), and \(\boldsymbol{\eta}\), \(\boldsymbol{\varepsilon}\), and \(\boldsymbol{\xi}\) are random vectors such that \(\boldsymbol{\varepsilon}\) is uncorrelated with \(\boldsymbol{\xi}\) (Neudecker \& Satorra, 1991). The parameters of the model are encapsulated in four matrices: \(\boldsymbol{\Lambda}\) contains the factor loadings, \(\boldsymbol{\Psi}\) contains the covariance matrix of \(\boldsymbol{\xi}\), \(\boldsymbol{B}_0\) contains the regression parameters of the structural model, and \(\boldsymbol{\Theta}\) contains the covariance matrix of \(\boldsymbol{\varepsilon}\).

As the covariance matrices are symmetric, we can construct the full parameter vector \(\boldsymbol{\delta}\) as follows:
\begin{equation}
\boldsymbol{\delta} = \left[ (\text{vec} \, \boldsymbol{\Lambda})^T, ( \text{vech} \, \boldsymbol{\Psi} )^T, (\text{vec} \, \boldsymbol{B}_0)^T, (\text{vech} \, \boldsymbol{\Theta})^T \right]^T
\label{eq:sempar}
\end{equation}
where the \(\text{vec}\) operator transforms a matrix into a vector by stacking the columns, and the \(\text{vech}\) operator does the same but eliminates the supradiagonal elements of the matrix. Specific models impose specific restrictions on this parameter vector. This leads to a subset of \emph{free parameters} \(\boldsymbol{\theta}\). \(\boldsymbol{\delta}\) is identified through these restrictions: \(\boldsymbol{\delta} = \boldsymbol{\delta}(\boldsymbol{\theta})\), and as a result, \(\boldsymbol{\delta}\) is estimated if \(\boldsymbol{\theta}\) is estimated.

\hypertarget{least-squares-and-maximum-likelihood-objectives}{%
\subsubsection{Least squares and maximum likelihood objectives}\label{least-squares-and-maximum-likelihood-objectives}}

Thus, \(\boldsymbol{\theta}\) contains all the free parameters. The model-implied covariance matrix \(\boldsymbol{\Sigma}(\boldsymbol{\theta})\) is a function of the free parameters, defined as follows (Bock \& Bargmann, 1966; Jöreskog, 1966):

\begin{equation}
\boldsymbol{\Sigma}(\boldsymbol{\theta}) = \boldsymbol{\Lambda}\boldsymbol{B}^{-1}\boldsymbol{\Psi}\boldsymbol{B}^{-T}\boldsymbol{\Lambda}^T + \boldsymbol{\Theta}
\label{eq:semsig}
\end{equation}

where \(\boldsymbol{B} = \boldsymbol{I} - \boldsymbol{B}_0\) is assumed to be non-singular.

In order to estimate \(\boldsymbol{\theta}\), an objective function needs to be defined. The objective in general is to reduce the distance between the model-implied covariance matrix \(\boldsymbol{\Sigma}(\boldsymbol{\theta})\) and the observed covariance matrix \(\boldsymbol{S}\): the model fits better if the model-implied covariance matrix more closely resembles the observed covariance matrix. The maximum-likelihood (ML) objective function \(F_{\text{ML}}\) is such a distance measure. Under the assumption that the observed covariance matrix follows a Wishart distribution or, equivalently, the observations \(\boldsymbol{z}\) follow a multivariate normal distribution, the maximum-likelihood fit function is the following (Bollen, 1989; Jöreskog, 1967):
\begin{equation}
F_{\text{ML}}(\boldsymbol{\theta}) = \log \left|\boldsymbol{\Sigma}(\boldsymbol{\theta})\right| + \text{tr}\left[\boldsymbol{S}\boldsymbol{\Sigma}^{-1}(\boldsymbol{\theta})\right]
\label{eq:semfit}
\end{equation}

Note that the ML fit function is a special case of the generalized least squares (GLS) fit function (Browne, 1974) which is defined as the following quadratic form:
\begin{equation}
F_{\text{GLS}}(\boldsymbol{\theta}) = (\boldsymbol{s} - \boldsymbol{\sigma}(\boldsymbol{\theta}))^T \boldsymbol{W} (\boldsymbol{s} - \boldsymbol{\sigma}(\boldsymbol{\theta}))
\end{equation}
Where \(\boldsymbol{s} = \text{vech} \, \boldsymbol{S}\), and \(\boldsymbol{\sigma}(\boldsymbol{\theta}) = \text{vech} \, \boldsymbol{\Sigma}(\boldsymbol{\theta})\). Here, \(F_{\text{GLS}} = F_{\text{ML}}\) when \(\boldsymbol{W} = 2^{-1}\boldsymbol{D}^T(\boldsymbol{\Sigma}^{-1}(\boldsymbol{\theta}) \otimes \boldsymbol{\Sigma}^{-1}(\boldsymbol{\theta}))\boldsymbol{D}\) (Neudecker \& Satorra, 1991), where \(\boldsymbol{D}\) is the duplication matrix and \(\otimes\) indicates the Kronecker product.

With this formulation, the gradient \(\boldsymbol{g}(\boldsymbol{\theta})\) of \(F_{\text{GLS}}\) with respect to the parameters \(\boldsymbol{\theta}\) and the Hessian \(\boldsymbol{H}(\boldsymbol{\theta})\) -- the matrix of second-order derivatives -- were derived by Neudecker \& Satorra (1991). These two quantities are the basis for standard errors, robust statistical tests for model fit (Satorra \& Bentler, 1988), as well as fast and reliable Newton-type estimation algorithms (Lee \& Jennrich, 1979). One such algorithm is the Newton-Raphson algorithm, where the parameter estimates at iteration \(i + 1\) are defined as the following function of the estimates at iteration \(i\):

\begin{equation}
\boldsymbol{\theta}^{(i + 1)} = \boldsymbol{\theta}^{(i)}-\boldsymbol{H}^{-1}(\boldsymbol{\theta}^{(i)})\cdot \boldsymbol{g}(\boldsymbol{\theta}^{(i)})
\label{eq:nr}
\end{equation}

Together, the objective function and the algorithm comprise an \emph{estimator} -- a way to compute parameter estimates using the data. The estimator described here is used extensively in software such as \texttt{lavaan} (Rosseel, 2012). Note that it is developed specifically for GLS estimation of SEM. With every extension to GLS, this work needs to be redone: a bespoke new estimator -- objective, gradient, Hessian, and algorithm -- needs to be derived and implemented.

\hypertarget{the-limits-of-ml-gls-estimation}{%
\subsubsection{The limits of ML / GLS estimation}\label{the-limits-of-ml-gls-estimation}}

There are several situations in which standard estimation methods lead to undesirable results in the form of nonconvergence or improper solutions (Chen, Bollen, Paxton, Curran, \& Kirby, 2001; Revilla \& Saris, 2013). Roughly, there are two ways in which this may happen: (a) a mismatch between the objective function and the algorithm used to find the optimum -- the ``optimizer'' -- and (b) a mismatch between the objective function and the information in the dataset. The former problem is a pure optimization problem. An example of an objective-optimizer mismatch is in mixture models. Here, the log-likelihood function may be nonconvex, meaning it has more than one local optimum. Newton-style algorithms such as Newton-Raphson may converge to local minima or saddle points, thus these models require more iterative estimation methods such as expectation maximization (EM).

The second problem relates to empirical underidentification (Rindskopf, 1984). Roughly, this means that the observed data contain little information about one or more parameters in the model. As a result, the model may fail to converge, or improper solutions such as negative estimated variances -- ``Heywood cases'' -- may occur. Empirical underidentification can be avoided through changing the objective function, by including prior information about parameters or partially constraining the parameter space. For example, some authors have suggested constraining variance parameters to be strictly positive (Helm, Castro-Schilo, \& Oravecz, 2017; Rindskopf, 1983). This changes the smoothness of the objective function, meaning that estimating such altered models requires specific parameterizations and estimation software because of the aforementioned objective-optimizer mismatch (Bentler \& Weeks, 1980; Muthén \& Asparouhov, 2012). Note that such restrictions also entail the dangers of introducing small-sample bias and obscuring potential misspecifications, which remains a separate theoretical issue (Chen et al., 2001; Kolenikov \& Bollen, 2012).

A good example of SEM reaching the limits of classical estimation is the recently introduced LASSO regularization (Jacobucci et al., 2016). The LASSO was designed for variable selection in high-dimensional data -- an empirically underidentified situation where there are few observations and many parameters. LASSO adds a nonsmooth penalty to the log-likelihood objective leading to shrinkage and selection of parameters (Hastie, Tibshirani, \& Wainwright, 2015). In order to compute the optimum of the nonsmooth LASSO objective function in regression, a specialized iterative algorithm needed to be introduced (Friedman et al., 2010) which does not rely on the Hessian or its inversion as Newton-type algorithms do. To perform such variable selection in structural equation models, this algorithm was ported to the SEM framework in a custom software package (\texttt{regsem}; Jacobucci, 2017), which also includes several other penalties.

The feasibility of such modifications to SEM can currently only be tested after custom software is written. We suggest a more general approach, allowing for extensions without rewriting the optimizer in a specialized package. Although this has been suggested before (Cudeck, Klebe, \& Henly, 1993), our approach expands the range of possible extensions using two key developments: (a) adaptive first-order optimizers and (b) computation graphs with automatic differentiation.

\hypertarget{adaptive-first-order-optimizers}{%
\subsection{Adaptive first-order optimizers}\label{adaptive-first-order-optimizers}}

We suggest using adaptive first-order optimizers to extend SEM beyond the existing estimation methods. Adaptive first-order optimizers are a class of optimization algorithms designed to work even under nonconvexity and nonsmoothness. Some early algorithms such as RMSProp (Tieleman \& Hinton, 2012) were originally developed with deep learning in mind, where nonconvexity, non-smoothness, and high-dimensional parameter spaces are common. Therefore, we consider these methods excellent candidates for estimating an expanding class of SEM models, as they have historically done for neural networks. The idea of using first-order optimizers for SEM is by no means new (Lee \& Jennrich, 1979), but the recent developments in this area have made it a feasible approach.

The simplest first-order optimizer is gradient descent, which uses the gradient \(\boldsymbol{g}(\boldsymbol{\theta})\) of the objective with respect to the parameters to guide the direction that each parameter should move towards. The gradient is combined with a step size \(s\) so that in each iteration \(i\) of gradient descent the parameters are moved a small amount towards the direction of the negative gradient evaluated at the current parameter values:

\begin{equation}
\boldsymbol{\theta}^{(i + 1)} = \boldsymbol{\theta}^{(i)} - s \cdot \boldsymbol{g}(\boldsymbol{\theta}^{(i)})
\label{eq:graddesc}
\end{equation}

This algorithm has a similar structure to the Newton-Raphson method shown in Equation \eqref{eq:nr}. In that algorithm, the step size \(s\) in each iteration is replaced by the inverse of the Hessian matrix. Gradient descent is thus a simplified version of the methods currently in use for optimizing SEM. Because it does not use the Hessian, it continues to function when the objective is not smooth or not convex. Computationally, it is also more tractable, foregoing the need to compute the full Hessian matrix. However, it is necessary to determine the correct step size \(s\). This is not a trivial problem: with an improperly tuned step size, the algorithm may never converge.

\begin{figure}[H]

{\centering \includegraphics[width=0.6\linewidth]{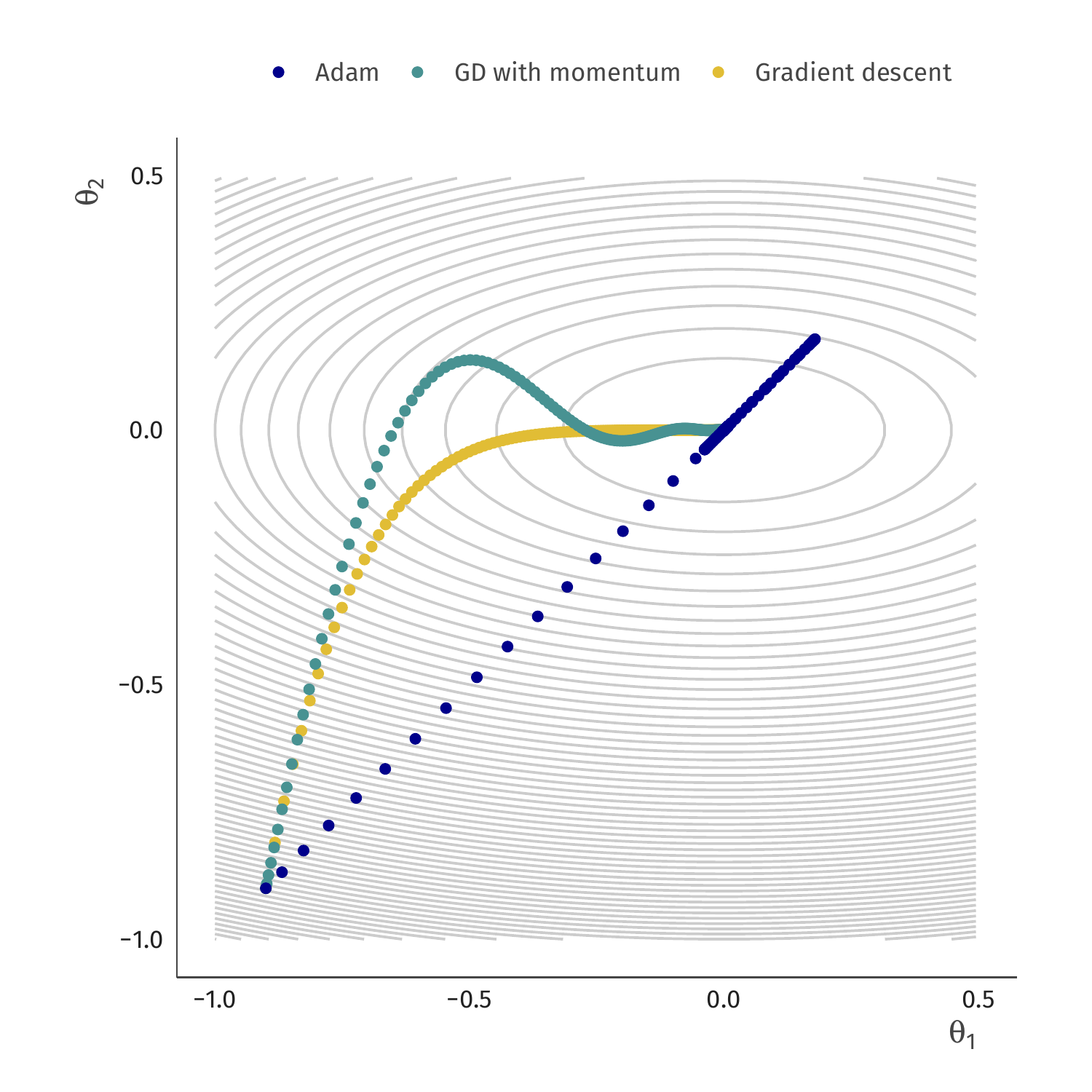} 

}

\caption{Three first-order algorithms finding the minimum of \(F(\boldsymbol{\theta}) = \theta_1^2 + 5\theta_2^2\) with starting value \(\boldsymbol{\hat{\theta}} =[-0.9, -0.9]\). Gradient descent uses the gradient and a fixed step size (\(s = 0.01\)) to update its parameter estimates. Gradient descent with momentum instead uses an exponential moving average of the gradients (decay of 0.9) with the same \(s\). Finally, Adam adds a moving average of the squared gradient (decay of 0.999) to adjust the step size per parameter, leading to a straight line to the minimum with an overshoot and return due to momentum. In this example, Adam converges fastest, and gradient descent is slowest.}\label{fig:adam}
\end{figure}

One of the state-of-the art adaptive first-order optimizers is Adam (Kingma \& Ba, 2014). It introduces two improvements to the framework of gradient descent (Figure \ref{fig:adam}). Firstly, it introduces momentum, where the direction in each iteration is not only the negative gradient of that iteration, but a \emph{moving average} of the entire history of gradients. Momentum allows Adam to move through local minima in the search for a global minimum by smoothing the path it takes in the parameter space. Secondly, Adam introduces a self-adjusting step size for each parameter, which is adjusted based on the \emph{variability} of the gradients over time: if the variability of the gradient of a parameter is smaller, Adam will take larger steps as it has more certainty about the direction the parameter should move in (and vice versa). This self-adjusting step size takes the place of computing and inverting the Hessian matrix. By using both the first and second moments of the history of the gradients, Adam is an adaptive optimizer capable of reliably optimizing a wide variety of objectives.

A relevant parallel to the development of adaptive first-order optimizers for deep learning is the recent advances in Bayesian SEM (Merkle \& Rosseel, 2015) and Bayesian posterior sampling in general. Here, too, the objective function may be nonconvex, e.g., in hierarchical models and with nonconjugate priors. Such objective functions may lead to inefficient behaviour for the Markov Chain Monte Carlo (MCMC) methods used to approximate posterior expectations. For this problem, Hamiltonial Monte Carlo (HMC) (Betancourt, 2017) has been developed, which introduces momentum in the proposal of a sample, thereby more efficiently exploring the posterior. This is the method implemented in Stan (Carpenter et al., 2017), which works for situations with many parameters and hyperparameters.

Adaptive first-order optimizers are one part of a pair of improvements that have enabled rapid growth of the deep learning field. The other is the development of computation graphs, an intuitive way of specifying the objective such that gradients can be computed automatically. Automatic gradient computation can enable a wide range of extensions to SEM without having to analytically derive the gradient and Hessian for each separate extension. In the next section, we explain the concept behind computation graphs and how they can be combined with optimizers such as Adam.

\hypertarget{computation-graphs-and-derivation-of-gradients}{%
\subsection{Computation graphs and derivation of gradients}\label{computation-graphs-and-derivation-of-gradients}}

A computation graph is a graphical representation of the operations required to compute a loss or objective value \(F(\boldsymbol{\theta})\) from (a vector of) parameters \(\boldsymbol{\theta}\) (Abadi et al., 2016). The full computation is split into a series of differentiable smaller computational steps. Each of these steps is represented as a node, with directed edges representing the flow of computation towards the final result. Because the nodes are differentiable, computing gradients of the final (or any intermediate) result with respect to any of its inputs is simple. Gradients are obtained by applying the chain rule of calculus starting at the node of interest and moving against the direction of the arrows in the graph. Thus, computation graphs are not only a convenient way of representing the outcome of a computation in a computer, but also immediately provide the derivatives, and, if required, second derivatives, that are necessary to optimize functions or estimate standard errors. For example, consider the ordinary least squares objective for linear regression:

\begin{equation}
F_{\text{LS}}(\boldsymbol{\beta}) = \sum_i(\boldsymbol{y}_i-\boldsymbol{x}_i\boldsymbol{\beta})^2 = (\boldsymbol{y} - \boldsymbol{X\beta})^T(\boldsymbol{y} - \boldsymbol{X\beta})
\label{eq:ls}
\end{equation}

\begin{figure}[H]

{\centering \includegraphics[width=1\linewidth]{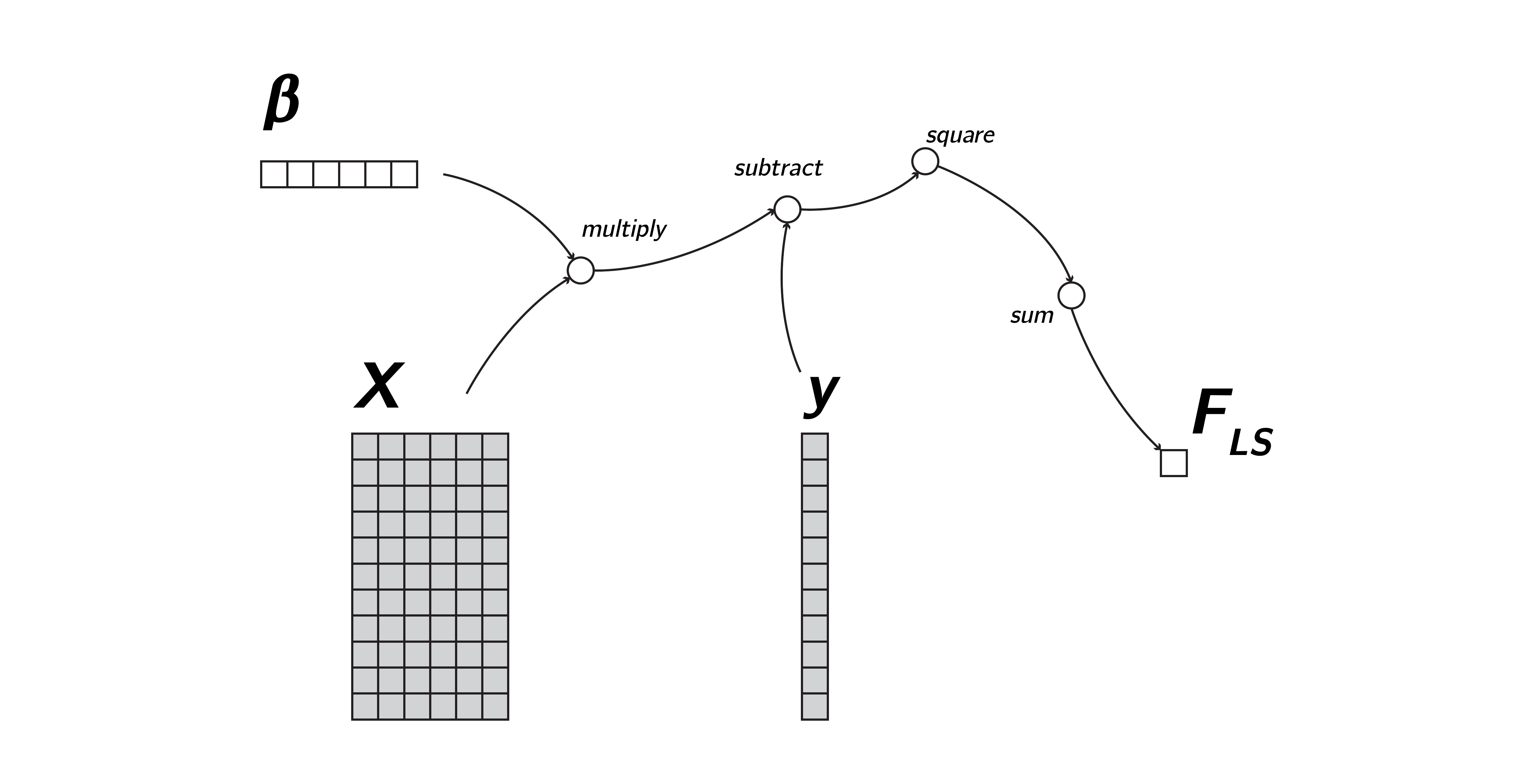} 

}

\caption{Least squares regression computation graph, mapping the regression coefficients (\(\boldsymbol{\beta}\)) to the least squares objective function \(F_{\text{LS}}\). The greyed-out parts contain elements which do not change as the parameters are updated, in this case observed data.}\label{fig:cg-ls}
\end{figure}

The computation graph of this objective function can be constructed as in Figure \ref{fig:cg-ls}. This figure represents the objective by ``unwrapping'' the equation from the inside outward into separate matrix operations: first, there is a matrix-vector multiplication of the design matrix \(\boldsymbol{X}\) with the parameter vector \(\boldsymbol{\beta}\). Then, the resulting \(n \times 1\) vector \(\boldsymbol{\hat{y}}\) is subtracted elementwise from the observed outcome \(\boldsymbol{y}\), and the result is squared, then summed to output a single squared error loss value \(F_{\text{LS}}\). The nodes in a computation graph may represent scalars, vectors, matrices, or even three- and higher-dimensional arrays. Generally, these nodes are referred to as \emph{tensors}.

Each of the operations in the graph has a registered derivative function. If it is known that the ``square'' operation \(f(x) = x^2\) is applied as in Figure \ref{fig:cg-ls}, the derivative \(f'(x)\) is \(2x\) and the second-order derivative \(f''(x)\) is \(2\). Thus, the gradient of the squared error tensor with respect to the residual tensor \(\boldsymbol{r}\) is \(2\boldsymbol{r}\).

Although automatic differentiation is an old idea (Wengert, 1964), its combination with state-of the art optimizers in software such as \texttt{Torch} (Collobert, Bengio, \& Mariéthoz, 2002; Paszke et al., 2017) and \texttt{TensorFlow} (Abadi et al., 2016) have paved the way for the current pace of deep learning research. Before the development and implementation of the computation graph, each neural network configuration (model) required specialized work on the part of the researchers who introduced it to provide a novel estimator. Thanks to computation graphs, researchers can design generic neural nets without needing to invent a bespoke estimator, a development that has greatly accelerated progress in this area. For SEM, we see a similar situation at the moment: each development or extension of the model currently requires a new algorithm that is capable of estimating its parameters. By applying computation graphs to SEM, we hope to greatly accelerate the process of developing novel SEM models. In the next section we combine the parameter configuration developed for SEM with the computation graphs and optimizers developed for deep learning to create a more flexible form of SEM.

\hypertarget{a-more-flexible-sem-using-computation-graphs}{%
\section{A more flexible SEM using computation graphs}\label{a-more-flexible-sem-using-computation-graphs}}

In this section, we develop the computation graph and parameter configuration to perform structural equation modeling using \texttt{TensorFlow} (Abadi et al., 2016). Then, we outline how this computation graph can be edited to extend SEM to novel situations, and how additional penalties can be imposed on any parameter in the model.

\hypertarget{the-sem-computation-graph}{%
\subsection{The SEM computation graph}\label{the-sem-computation-graph}}

\begin{figure}[H]

{\centering \includegraphics[width=1\linewidth]{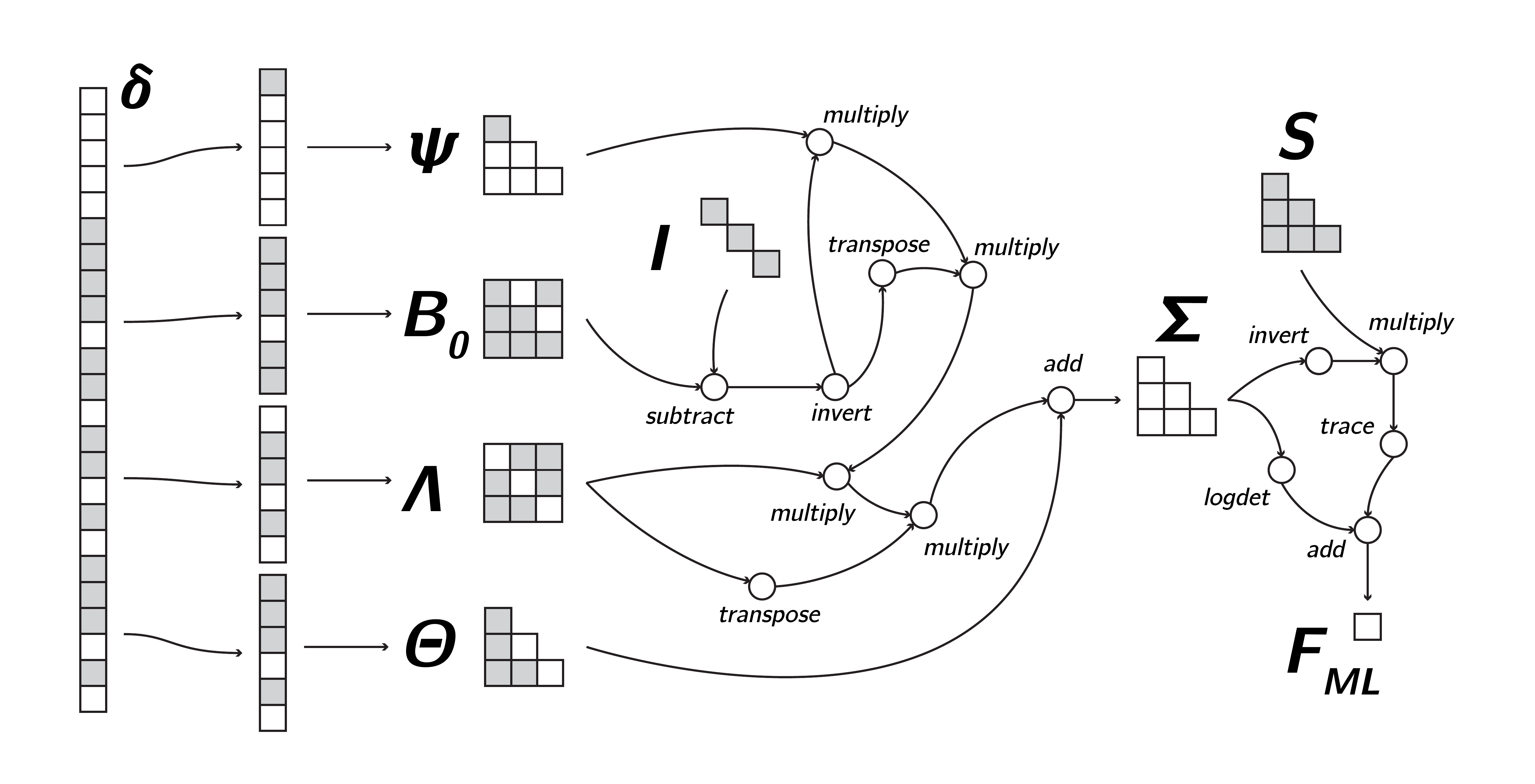} 

}

\caption{Full SEM computation graph, mapping the parameters (\(\boldsymbol{\delta}\)) to the maximum likelihood fit function \(F_{\text{ML}}\). The greyed-out parts contain elements which do not change as the parameters are updated, meaning either observed data or constrained parameters. (NB: The constrained elements in this graph are not representative of a specific model).}\label{fig:cg-sem}
\end{figure}

The SEM computation graph was constructed using the \texttt{tensorflow} R package (Allaire \& Tang, 2019). It is displayed in Figure \ref{fig:cg-sem}. From left to right, a parameter vector \(\boldsymbol{\delta}\) is first instantiated with constrained elements, such that the free parameters represent \(\boldsymbol{\theta}\). Then, this vector is split into the separate vectors as in Equation \eqref{eq:sempar}. These vectors are then reshaped into the four SEM all-y matrices, using a duplication matrix for the symmetric matrices \(\boldsymbol{\Psi}\) and \(\boldsymbol{\Theta}\).

In the next part, these matrices are transformed to the model-implied covariance matrix \(\boldsymbol{\Sigma}(\boldsymbol{\theta})\) by unwrapping Equation \eqref{eq:semfit} from the inside outward: \(\boldsymbol{B}^{-1}\) is constructed as \((\boldsymbol{I} - \boldsymbol{B}_0)^{-1}\), then \(\boldsymbol{\Psi}\) is premultiplied by this tensor and postmultiplied by its transpose. Then, the resulting tensor itself is pre - and postmultiplied by \(\boldsymbol{\Lambda}\) and \(\boldsymbol{\Lambda}^T\), respectively. Lastly, \(\boldsymbol{\Theta}\) is added to construct the implied covariance tensor.

The last part is the graphical representation of the ML fit function from Equation \eqref{eq:semfit}. \(\boldsymbol{\Sigma}(\boldsymbol{\theta})\) is inverted, then premultiplied by \(\boldsymbol{S}\), and the trace of this tensor is added to the log determinant of the inverse of \(\boldsymbol{\Sigma}(\boldsymbol{\theta})\). The resulting tensor, a scalar value, is the \(F_{\text{ML}}(\boldsymbol{\theta})\) objective function for SEM.

Each operation in Figure \ref{fig:cg-sem} carries with it information about its gradient. \texttt{TensorFlow} can therefore automatically convert the SEM computation graph into gradients with respect to the model parameters. The Hessian can also be obtained automatically by applying the same principle to the gradients. Note that these correspond to the observed score and information matrix, rather than their expected versions derived under the null hypothesis of model correctness. \texttt{TensorFlow} also provides facilities for easily optimizing computation graphs using these quantities. The flexibility of the computation graph notation lies in the ease with which sections of the graph can be edited to suit different needs. Below we demonstrate how this formulation can be leveraged to estimate nonstandard SEM models.

\hypertarget{different-objective-functions}{%
\subsection{Different objective functions}\label{different-objective-functions}}

The computation graph approach allows completely different objective functions to be implemented with relative ease. One such objective was coined by Siemsen \& Bollen (2007), who introduce least absolute deviation (LAD) estimation. Their motivation is the performance of the LAD estimator as a robust estimation method in other fields. Note that while Siemsen and Bollen find limited relevance for this SEM estimator in terms of performance, we consider it to be an excellent showcase of the flexibility of our approach. This objective does not fit in the GLS approach of Browne (1974). The LAD estimator implies the following objective:

\begin{equation}
F_{\text{LAD}}(\boldsymbol{\theta}) = \sum_{i,j} \left| \boldsymbol{\Sigma}(\boldsymbol{\theta})_{i,j}- \boldsymbol{S}_{i, j} \right|
\label{eq:semlad}
\end{equation}

A computational advantage of this objective relative to the GLS fit function is that there is no need to invert \(\boldsymbol{\Sigma}(\boldsymbol{\theta})\). The work of Siemsen \& Bollen (2007) focuses on developing a greedy genetic evolution numerical estimation algorithm which performs a search over the parameter space. Using this optimization algorithm, they show that the LAD estimator may outperform the ML estimator in very specific situations.

\begin{figure}[H]

{\centering \includegraphics[width=1\linewidth]{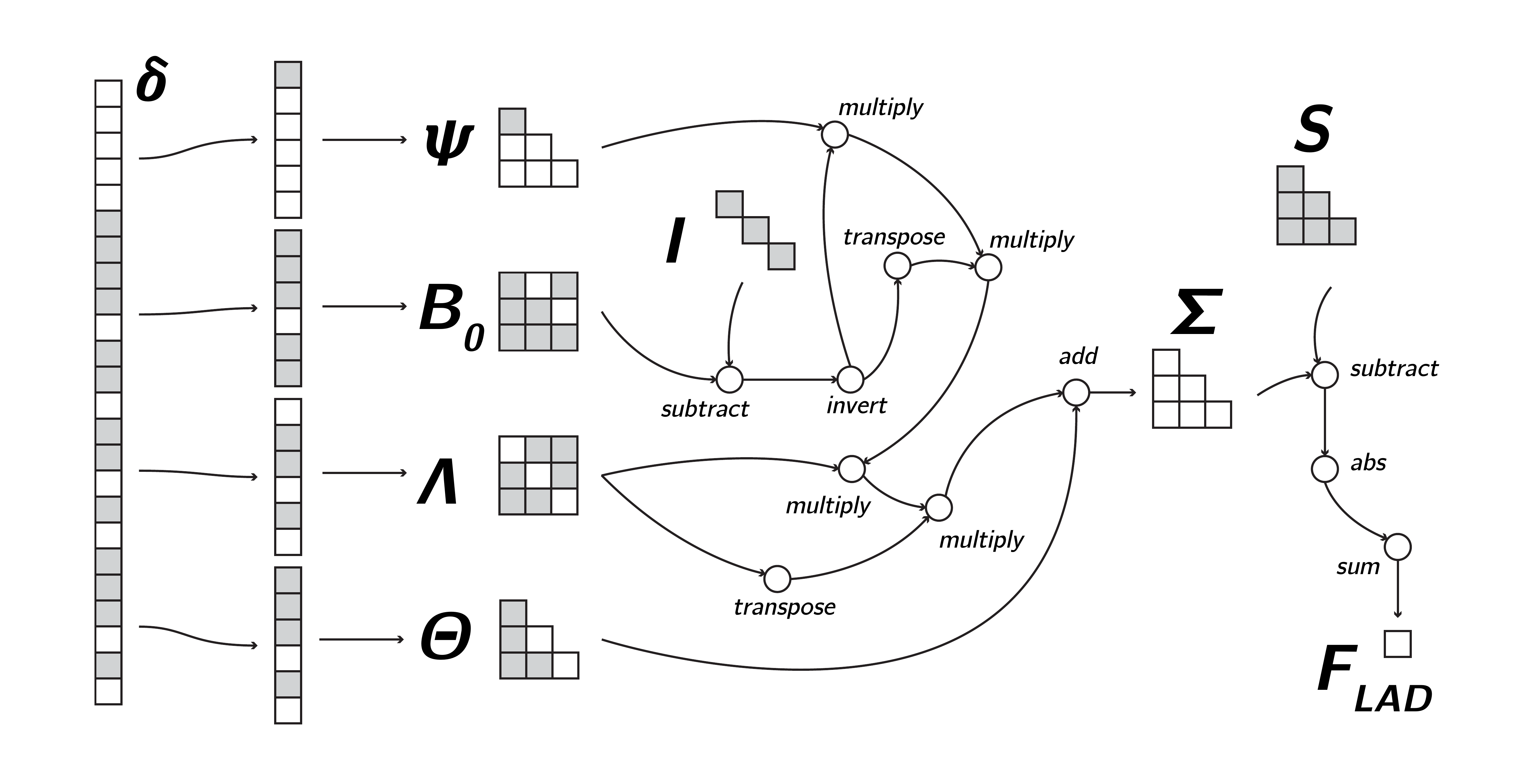} 

}

\caption{Full SEM computation graph for the least absolute deviation (LAD) objective. Compared to the ML fit function, the last part of the graph contains different operations.}\label{fig:cg-lad}
\end{figure}

Constructing the LAD estimator in the computation graph framework means replacing the ML fit operations with the LAD operations. This is shown in Figure \ref{fig:cg-lad}. Note that compared to the ML objective, there are fewer operations, and the inversion operation of the implied covariance matrix is removed. This change is trivial to make given the SEM computation graph, and we show in Section \ref{sec:ex-lad} that such alternative objective functions can be estimated with the Adam optimizer.

\hypertarget{regularization-through-parameter-penalization}{%
\subsection{Regularization through parameter penalization}\label{regularization-through-parameter-penalization}}

A more useful modification that can be readily made using the SEM computation graph is the addition of penalties to the parameters (Jacobucci et al., 2016). Such penalties \emph{regularize} the model, which may prevent overfitting and improve generalizability (Hastie et al., 2015). There is a wide variety of parameter penalization procedures, but the most common methods are ridge and LASSO. In regression, the widely used elastic net (Zou \& Hastie, 2005) is a combination of the LASSO and ridge penalties. The objective function for elastic net is the following:

\begin{equation}
F_{\text{EN}}(\boldsymbol{\beta}) = F_{\text{LS}}(\boldsymbol{\beta}) + \lambda_1 \left\lVert \boldsymbol{\beta} \right\rVert_1 + \lambda_2 \left\lVert \boldsymbol{\beta} \right\rVert_1^2
\label{eq:elanet}
\end{equation}

where \(\left\lVert \boldsymbol{\beta} \right\rVert_1 = \sum_p|\beta_p|\), and \(\lambda_1\) and \(\lambda_2\) are hyperparameters which determine the amount of LASSO and ridge shrinkage, respectively. By setting \(\lambda_1\) to zero we obtain L2 (ridge) shrinkage, and setting \(\lambda_2\) to zero yields the L1 (LASSO). Nonzero values for both parameters combines the two approaches, which has been shown to encourage a grouping effect in regression, where strongly correlated predictors tend to be in or out of the model together (Zou \& Hastie, 2005).

Friedman et al. (2010) have developed an efficient algorithm for estimating the elastic net for generalized linear models and have implemented this in their package \texttt{glmnet}. For SEM, Jacobucci et al. (2016) have created a package for performing penalization by adding the elastic net penalty to the ML fit function. Their implementation uses the RAM notation (McArdle \& McDonald, 1984), and their suggestion is to penalize either the \(\boldsymbol{A}\) matrix (factor loadings and regression coefficients), or the \(\boldsymbol{S}\) matrix (residual covariances).

In the field of deep learning, parameter penalization is one of the key mechanisms by which massively overparameterized neural networks are estimated (Goodfellow et al., 2016). Regularization is therefore a core component of various software libraries for deep learning, including \texttt{TensorFlow}. The optimizers implemented in these libraries, such as Adam, are tried and tested methods for estimation of neural networks with penalized parameters, which is an active field of research (e.g., Scardapane, Comminiello, Hussain, \& Uncini, 2017).

\begin{figure}[H]

{\centering \includegraphics[width=1\linewidth]{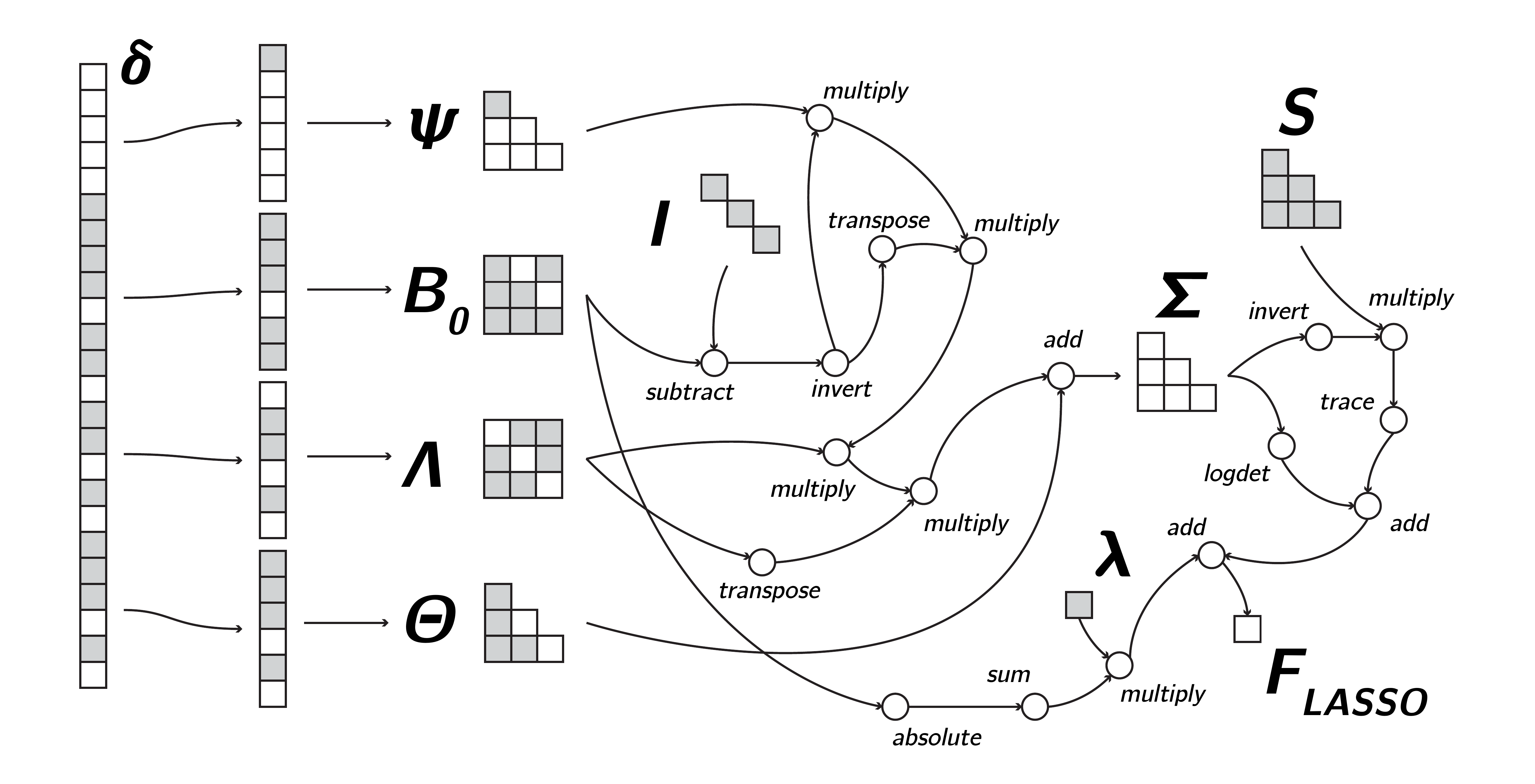} 

}

\caption{\(\boldsymbol{B}_0\) LASSO computation graph with a pre-defined \(\lambda\) tuning parameter.}\label{fig:cg-lasso}
\end{figure}

In the SEM computation graph, the LASSO penalty on the regression parameters can be implemented by adding a few nodes to the ML fit graph. This is displayed in Figure \ref{fig:cg-lasso}. The absolute value of the elements of the \(\boldsymbol{B}_0\) tensor are summed, and the resulting scalar is multiplied by the tuning parameter. The resulting value is then added to the maximum likelihood fit tensor to construct the lasso objective \(F_{\text{LASSO}}(\boldsymbol{\theta})\).

Ridge penalties for the \(\boldsymbol{B}_0\) matrix can be implemented in similar fashion, but instead of an ``absolute value'' operation, the first added node is a ``square'' operation. These penalties can be added to any tensor in the computation graph, meaning penalization of the factor loadings or the residual covariances, or even a penalty on \(\boldsymbol{B}\) is quickly implemented. The elastic net penalty specifically can be implemented by imposing both a ridge and a lasso penalty on the tensor of interest.

Note that each additional penalty comes with its own parameter to be selected -- a process called ``hyperparameter tuning''. Tuning of penalty parameters is traditionally done through cross-validation; \texttt{glmnet} (Friedman et al., 2010) provides a function for automatically selecting the penalization strength in regression models through this method. Another method is through inspecting model fit criteria. For example, Jacobucci et al. (2016) suggest selecting the penalty parameter through the BIC or the RMSEA, where the degrees of freedom is determined by the amount of \emph{nonzero} parameters, which changes as a function of the penalization strength. Another example is penalized network estimation, where Epskamp, Borsboom, \& Fried (2018) suggest hyperparameter tuning through an extended version of the BIC. In Bayesian optimization, there is another option. Here, van Erp, Oberski, \& Mulder (2019) show that a prior can be set on the penalty parameter -- a ``hyperprior'' -- and in this way the parameter itself is learned along with the model: the ``full Bayes'' approach. In the deep learning literature, this is called Bayesian optimization or gradient-based optimization of hyperparameters (Bengio, 2000).

In the field of variable selection, various implementations of the LASSO have been shown to exhibit desirable properties (e.g., Van De Geer, Bühlmann, \& others, 2009), making it a natural choice for obtaining sparsity in structural equation models. In addition, there are extensions which decrease bias in high-dimensional situations (Fan \& Li, 2001; Zhang, 2010). In the Bayesian case, the spike-and-slab prior is such an extension which shows less bias when compared to the LASSO-equivalent Laplace prior (van Erp et al., 2019). Ročková \& George (2018) translated the fundamentally probabilistic Bayesian spike-and-slab prior into a penalized likelihood objective. With the SEM computation graph, we can implement something similar, combining a Laplace spike with a normal slab by using a mixture of LASSO and ridge penalties:

\begin{equation}
F_{\text{SS}}(\boldsymbol{\theta}, \lambda_1, \lambda_2, \pi) = F_{\text{ML}}(\boldsymbol{\theta}) + \pi \cdot \lambda_1 \left\lVert \boldsymbol{\theta} \right\rVert_1 + (1 - \pi) \cdot \lambda_2 \left\lVert \boldsymbol{\theta} \right\rVert_1^2
\label{eq:spikeslab}
\end{equation}

\begin{figure}[H]

{\centering \includegraphics[width=1\linewidth]{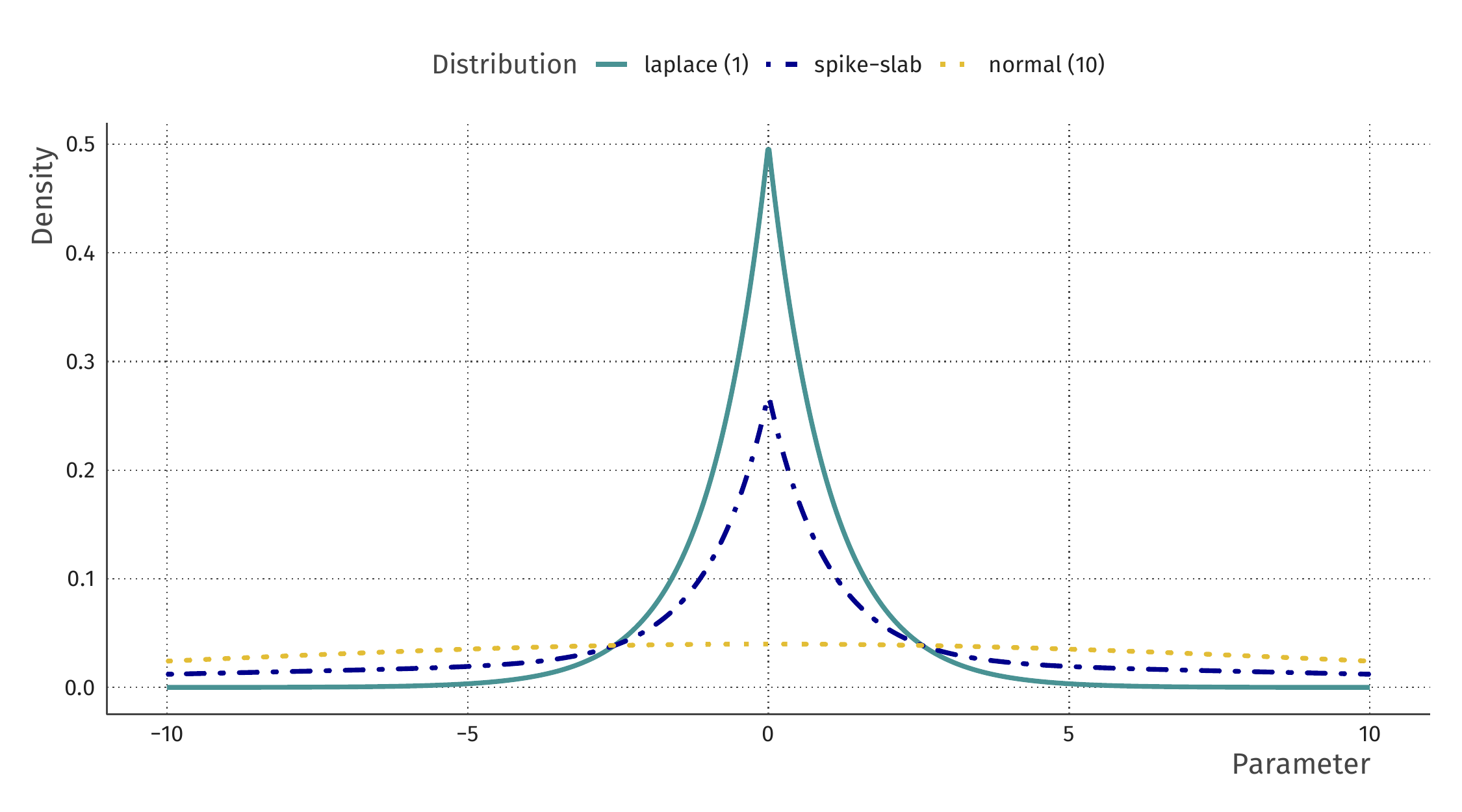} 

}

\caption{Prior induced through LASSO, ridge, and mixture penalties. The mixture, borrowing from the Bayesian ideas of a spike-and-slab prior, has larger tails than the LASSO, leading to less bias due to less shrinkage in nonzero parameters. The numbers in the parentheses indicate dispersion.}\label{fig:ex-sslab}
\end{figure}

The resulting equivalent prior is shown in Figure \ref{fig:ex-sslab}. The spike-and-slab prior we have implemented is a mixture of a Laplace prior (spike) and a wide normal prior (slab). The mixing parameter \(\pi\) encodes prior knowledge of sparsity: what proportion of the parameters should be in the spike. Due to the slab component, it has larger tails than the Laplace prior, leading it to shrink large parameter values less. These larger tails lead this prior to exhibit less bias in high-dimensional situations. This penalty is novel to SEM, and poses challenges to traditional estimation methods due to its nonconvexity, but it follows naturally from the computation graph approach. In Section \ref{sec:regfa}, we show how this penalty compares to the LASSO and ML fit functions.

\hypertarget{standard-errors-and-model-tests}{%
\subsection{Standard errors and model tests}\label{standard-errors-and-model-tests}}

In SEM, standard errors can be calculated through the Fisher information method, requiring only the Hessian of the log-likelihood at the maximum and the assumption that the distribution this log-likelihood is based on (usually Normal-theory) is correct. Additionally, the distributional assumption can be relaxed by using sandwich estimators, in SEM known as Satorra-Bentler (robust) standard errors. These need both the Hessian and the \(N \times P\) outer product matrix \(\Delta\) -- the case-wise first derivatives of the parameters w.r.t. the implied covariances \(\sigma(\theta)\) (Savalei, 2014). Sandwich estimators also lead to robust test statistics which are not sensitive to deviations from normality. In econometrics, many variations of the sandwich estimator are available, depending on whether the expected or observed information matrix is used (Kolenikov \& Bollen, 2012).

Computation graphs as outlined in this section are a general approach for obtaining parameter estimates of structural equation models. Moreover, for the ML computation graph (Figure \ref{fig:cg-sem}) it is also possible to obtain accurate standard errors because the observed information matrix -- the inverse of the Hessian of the log-likelihood -- is available automatically through the gradient computation in TensorFlow. In addition, through the same computation graph but with case-wise entering of the data, the outer product matrix \(\Delta\) can also be made available. Because these are the observed versions, computation of empirical sandwich (Huber-White) standard errors is possible. Naturally, an established alternative to these procedures is to bootstrap in order to obtain standard errors. Furthermore, the log-likelihood itself is directly available, thus information criteria such as AIC, BIC and SSABIC (Sclove, 1987), as well as normal-theory and robust test statistics (Satorra \& Bentler, 1988) can be computed more or less ``as usual''.

In principle, therefore, standard errors and test statistics are available when using the computation graph approach. However, in practice the computation graph can be edited arbitrarily by introducing penalties or a different fit function. In this case, no general guarantees can be given about the accuracy of the standard errors, the coverage probability of the confidence interval, or the asymptotic behaviour of model fit metrics derived from the obtained model. This is inherent to the flexibility of the computation graph approach: for existing methods in SEM, simulations have shown the performance of the current standard error solutions (including the bootstrap), but as extensions are introduced these results do not necessarily hold. For some extensions, there will be no adequate approximation to the standard error with accurate frequentist properties. For example, there is a large body of literature on standard error approximations for \(L1\) penalization (e.g., Fan \& Li, 2001), but the problem of obtaining penalized model standard errors is fundamentally unsolvable due to the bias introduced by altering the objective function away from the log-likelihood (Goeman, Meijer, \& Chaturvedi, 2018, p. 18). Not even the bootstrap can provide consistent standard error estimates in these situations (Kyung, Gill, Ghosh, Casella, \& others, 2010). Hence, software implementations of penalized regression (e.g., \texttt{glmnet}) consciously omit standard errors.

In situations beyond ML, our advice is to pay attention to the behaviour of existing fit criteria and standard errors. Using simulations for each new model and data case, the frequentist properties of the empirical confidence interval can be assessed and the type-I and type-2 errors of the (Satorra-Bentler) \(\chi^2\) test can be found. Those values can then be used to adjust the interpretation of the results in the analysis of the real data. If existing standard error approaches fail altogether, a viable solution may be to completely omit standard errors -- just as in the \(L1\) regression approach.

Note that all of the above holds similarly for Bayesian estimation, where the choice of prior influences the frequentist properties of the posterior, such as the credible interval coverage probability. Just as it is possible with the computation graph approach to create a nonconverging model with bad asymptotic behaviour, it is possible with Bayesian methods to create such a problematic model through the choice of nonsensical priors. Solutions in this case are also based on simulation, e.g., prior predictive checking (Gabry, Simpson, Vehtari, Betancourt, \& Gelman, 2019) or leave-one-out cross-validation (Vehtari, Gelman, \& Gabry, 2017).

In the next section, we show through a set of examples motivated by existing literature how our implementation of the SEM computation graph can be used to estimate regular structural equation models and their standard errors, as well as the extensions we have introduced in this section.

\hypertarget{examples}{%
\section{Examples}\label{examples}}

All the computation graphs and estimation methods described in this paper are implemented in the \texttt{computationgraph} branch of \texttt{tensorsem} (van Kesteren, 2019), available at \href{https://github.com/vankesteren/tensorsem}{github.com/vankesteren/tensorsem}. In this section, we use this software to show how our extensions compare to reference implementations, if available. The first example demonstrates that our approach gives the same estimates are obtained as \texttt{lavaan} (Rosseel, 2012) for a standard SEM with latent variables. We then demonstrate how non-standard extensions can be implemented with similar ease. We show how the LAD estimator yields similar parameters to the ML estimator in a factor analysis. Then we perform LASSO regression with observed and latent (MIMIC) outcomes, comparing our method to both \texttt{regsem} (Jacobucci et al., 2016) and the benchmark \texttt{glmnet} (Friedman et al., 2010). Lastly, we show how LASSO and spike-and-slab regularization can induce sparsity in the loadings of a confirmatory factor analysis (CFA).

Fully commented, reproducible code to create the results, plots, and tables in this section is available as supplementary material to this paper at the following link: \href{https://github.com/vankesteren/sem-computationgraphs}{github.com/vankesteren/sem-computationgraphs}.

\hypertarget{sec:ex-sem}{%
\subsection{Structural equation modeling}\label{sec:ex-sem}}

\begin{figure}[H]

{\centering \includegraphics[width=1\linewidth]{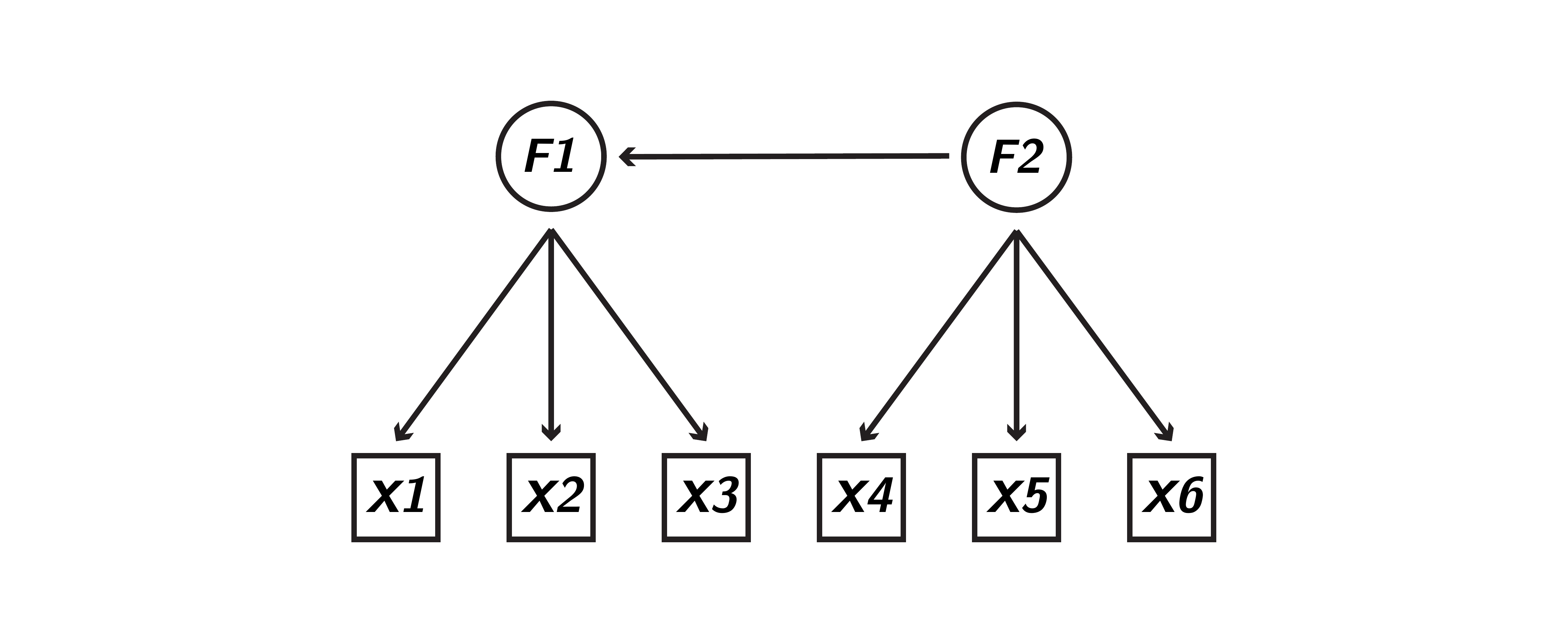} 

}

\caption{Structural equation model with two latent variables and six indicators used for comparing the results of \texttt{tensorsem} to \texttt{lavaan}. Indicators are variables from the Holzinger \& Swineford (1939) dataset. For clarity, residual variances have been omitted from the image.}\label{fig:ex-sem}
\end{figure}

To validate the computation graph approach and the Adam estimator for structural equation modeling with latent variables, we estimated a model with two latent variables and a structural coefficient as in Figure \ref{fig:ex-sem}. The indicators come from the data of Holzinger \& Swineford (1939) included in \texttt{lavaan}.

The \texttt{tensorsem} and \texttt{lavaan} packages both use the same model specification language developed by Rosseel (2012). Model estimation in \texttt{lavaan} happens through a Fisher scoring algorithm with a ML fit function, and \texttt{tensorsem} uses the Adam optimizer and the computation graph in Figure \ref{fig:cg-sem}.

\begin{figure}[H]

{\centering \includegraphics[width=1\linewidth]{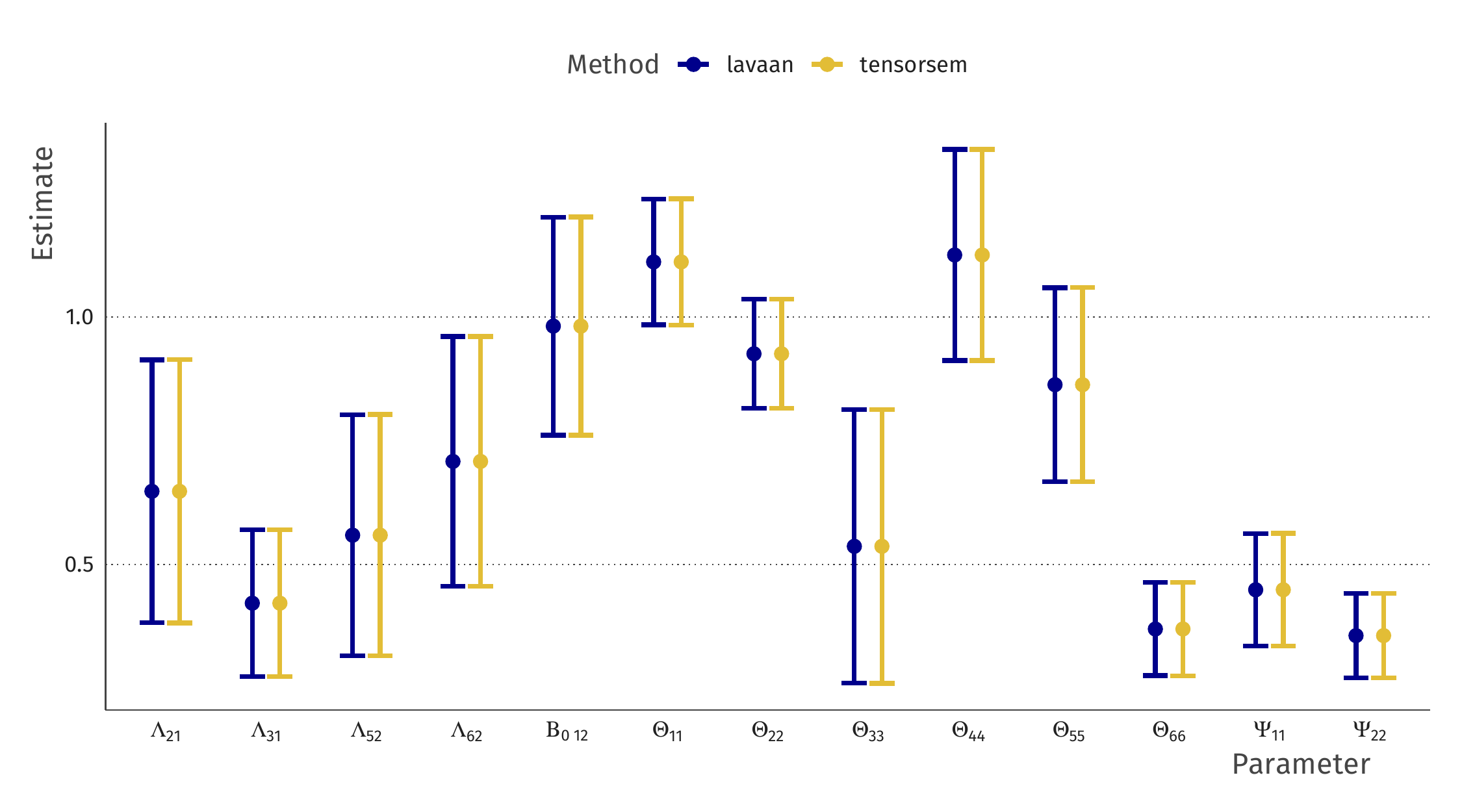} 

}

\caption{Parameter estimates and confidence intervals for a model with two latent variables and a structural parameter, comparing \texttt{lavaan} to \texttt{tensorsem}. Confidence intervals are constructed as \(\boldsymbol{\hat{\theta}}_i \pm 1.96 \times SE_i\), where the standard errors are computed from the observed information matrix.}\label{fig:res-sem}
\end{figure}

The results are shown in Figure \ref{fig:res-sem}. From the graph, it becomes clear that both the packages arrive at the same solution and, furthermore, the standard errors are the same across the different packages. In \texttt{lavaan} we use the observed information matrix for the standard errors. The standard errors in \texttt{tensorsem} are computed by taking the Hessian of the objective with respect to the free parameters, which is directly available from the computation graph and thus can be returned by the \texttt{TensorFlow} backend. The Hessian then needs to be inverted and rescaled (Bollen, 1989, p. 135) to obtain a consistent estimate \(\text{ACOV}(\boldsymbol{\theta})\) of the covariance matrix of the parameters under the assumption of multivariate normality and standard ML theory. The standard errors are the square root of the diagonal elements of this matrix. This standard error procedure comes from ML theory, meaning it is valid only for the ML objective function and under the assumption of multivariate normality. Distribution-robust standard error estimators could also be implemented straightforwardly, but this is outside the scope of the current paper.

From this example, we conclude that computation graphs and Adam optimization are capable of estimating structural equation models. In addition, as the solution obtained by \texttt{tensorsem} is the same as with other packages, it is possible compute the value of the log-likelihood objective function and its derivative fit measures such as \(\chi^2\), AIC, and BIC. In the next examples, we leverage the flexibility of this framework to extend SEM to novel situations.

\hypertarget{sec:ex-lad}{%
\subsection{LAD estimation}\label{sec:ex-lad}}

Although LAD estimation was shown to be beneficial only in very specific situations (Siemsen \& Bollen, 2007), it is an excellent showcase for the flexibility of the computation graph approach. Because the software developed by Siemsen \& Bollen (2007) is not available, we instead compare the LAD estimates to the ML estimates. In this case, there is no direct comparison method for the extension we introduce in \texttt{tensorsem}. The \texttt{tensorsem} LAD estimator is thus a novel way of estimating SEM.

\begin{figure}[H]

{\centering \includegraphics[width=1\linewidth]{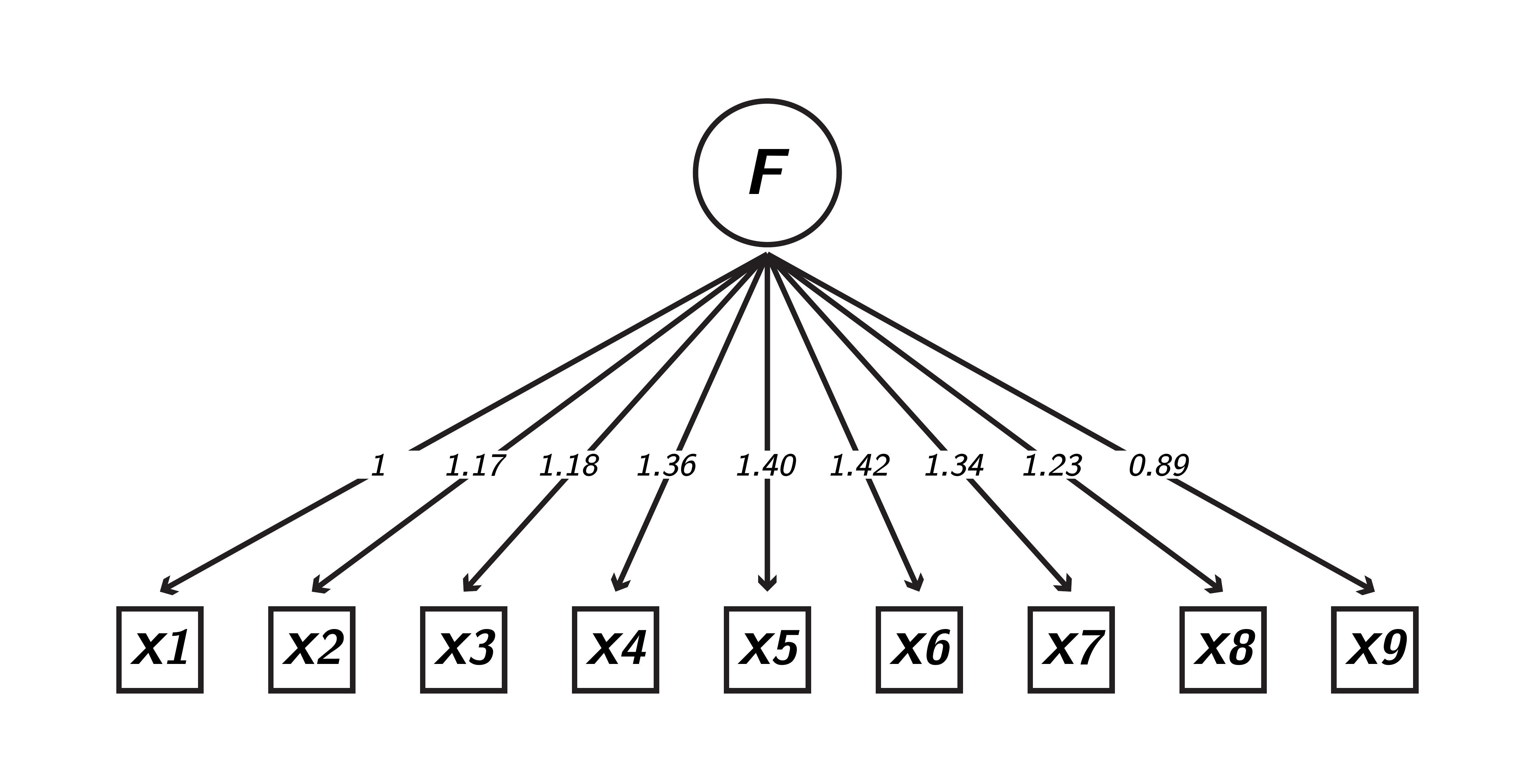} 

}

\caption{Factor analysis model used to generate data for comparing the least absolute deviation (LAD) estimator to the maximum likelihood (ML) estimator in \texttt{tensorsem}. Residual variances of the indicators were all set to 1.}\label{fig:ex-lad}
\end{figure}

For this example, we generate data of sample size 1000 from a one-factor model. For this data, we constrain the observed covariance matrix to the covariance matrix implied by the population model in Figure \ref{fig:ex-lad}. Since LAD estimation should be robust to outliers in the observed covariance matrix, which can happen in the trivial case of mistranscribing a covariance matrix into software, we also performed this on data with a ``contaminated'' covariance matrix: \(COV(X_1, X_3) = 2,\, COV(X_2, X_4) = 0.35\).

The results are shown in Table \ref{tab:res-lad}. The for ML estimation in \texttt{lavaan} and \texttt{tensorsem} again agree. With uncontaminated covariance matrix, the LAD estimates reach the same conclusion as the ML estimates. Note that although unbiased, LAD is relatively less efficient, but this effect is not visible with a sample size of 1000 for this model. With contamination in the covariance matrix, the LAD method shows no bias in this example, whereas the ML method does. Because the Hessian for the LAD objective is not invertible, the standard errors are not available using the \(\text{ACOV}(\boldsymbol{\theta})\) method described in section \ref{sec:ex-sem}. Siemsen \& Bollen (2007) solve this problem by bootstrapping, which is possible but outside the scope of the current paper.

\begin{table}[H]

\caption{\label{tab:res-lad}Parameter estimates comparing ML estimates to LAD estimates using both uncontaminated and contaminated covariance matrices. LAD is robust to the contamination of the covariance matrix.}
\centering
\begin{tabular}[t]{llcccccccc}
\toprule
  & X1 & X2 & X3 & X4 & X5 & X6 & X7 & X8 & X9\\
\midrule
lavaan (ML) & 1.00 & 1.17 & 1.18 & 1.36 & 1.40 & 1.42 & 1.34 & 1.23 & 0.89\\
tensorsem (ML) & 1.00 & 1.17 & 1.18 & 1.36 & 1.40 & 1.42 & 1.34 & 1.23 & 0.89\\
tensorsem (LAD) & 1.00 & 1.17 & 1.18 & 1.36 & 1.40 & 1.42 & 1.34 & 1.23 & 0.89\\
Contaminated (ML) & 1.13 & 1.05 & 1.30 & 1.25 & 1.38 & 1.40 & 1.32 & 1.21 & 0.88\\
Contaminated (LAD) & 1.00 & 1.17 & 1.18 & 1.36 & 1.40 & 1.42 & 1.34 & 1.23 & 0.89\\
\bottomrule
\end{tabular}
\end{table}

\begin{figure}[H]

{\centering \includegraphics[width=1\linewidth]{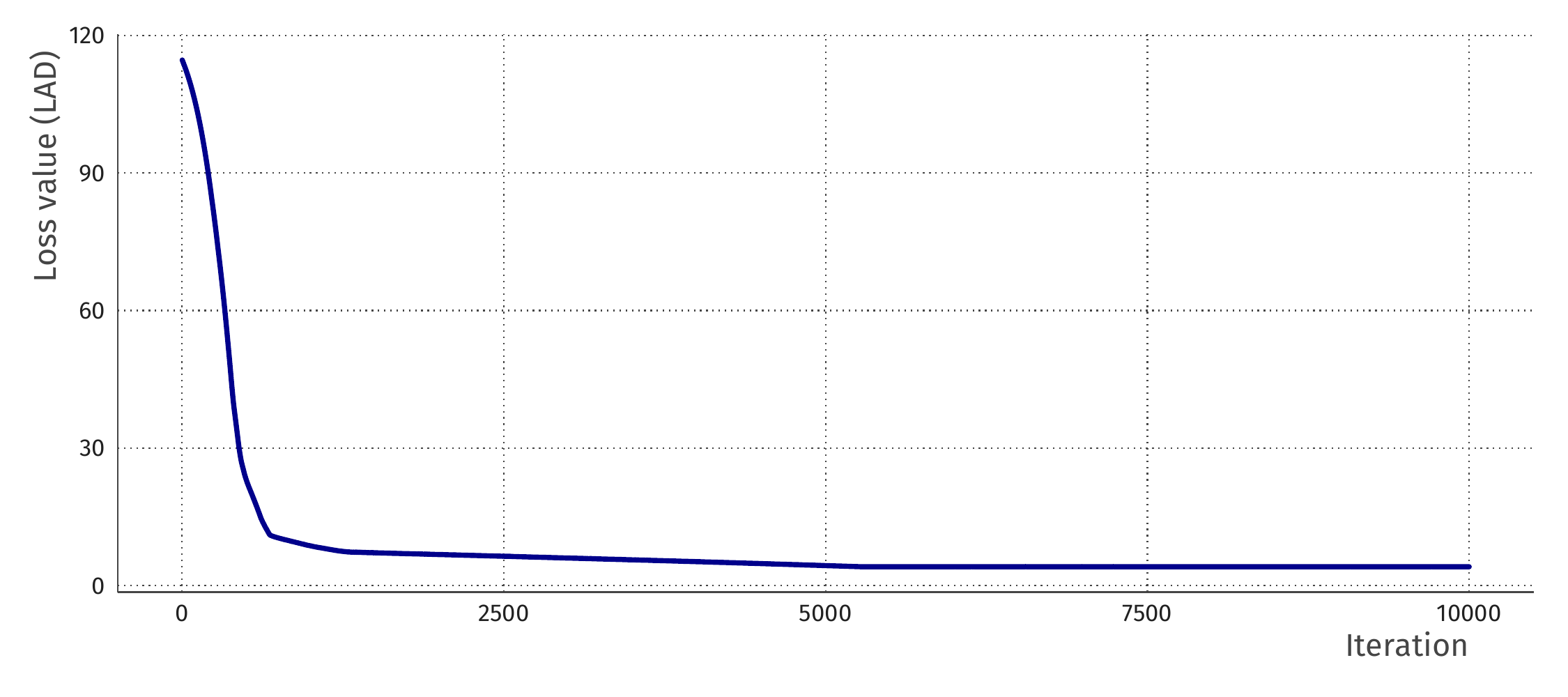} 

}

\caption{Plot of the value of the objective function over iterations of the LAD estimation procedure with a contaminated covariance matrix. As the optimizer iterates, the loss value decreases until convergence.}\label{fig:ex-lad-conv}
\end{figure}

The results from this example show that the objective function in \texttt{tensorsem} can be edited and that Adam still converges to a stable solution with this adjusted objective (Figure \ref{fig:ex-lad-conv}). The parameter estimates from LAD estimation approximate those obtained from ML estimation in a one-factor model with 9 indicators and 1000 samples. In addition, we have shown that LAD estimation is robust to contamination of the covariance matrix in the contrived example of this section.

\hypertarget{regularized-regression}{%
\subsection{Regularized regression}\label{regularized-regression}}

\begin{figure}[H]

{\centering \includegraphics[width=1\linewidth]{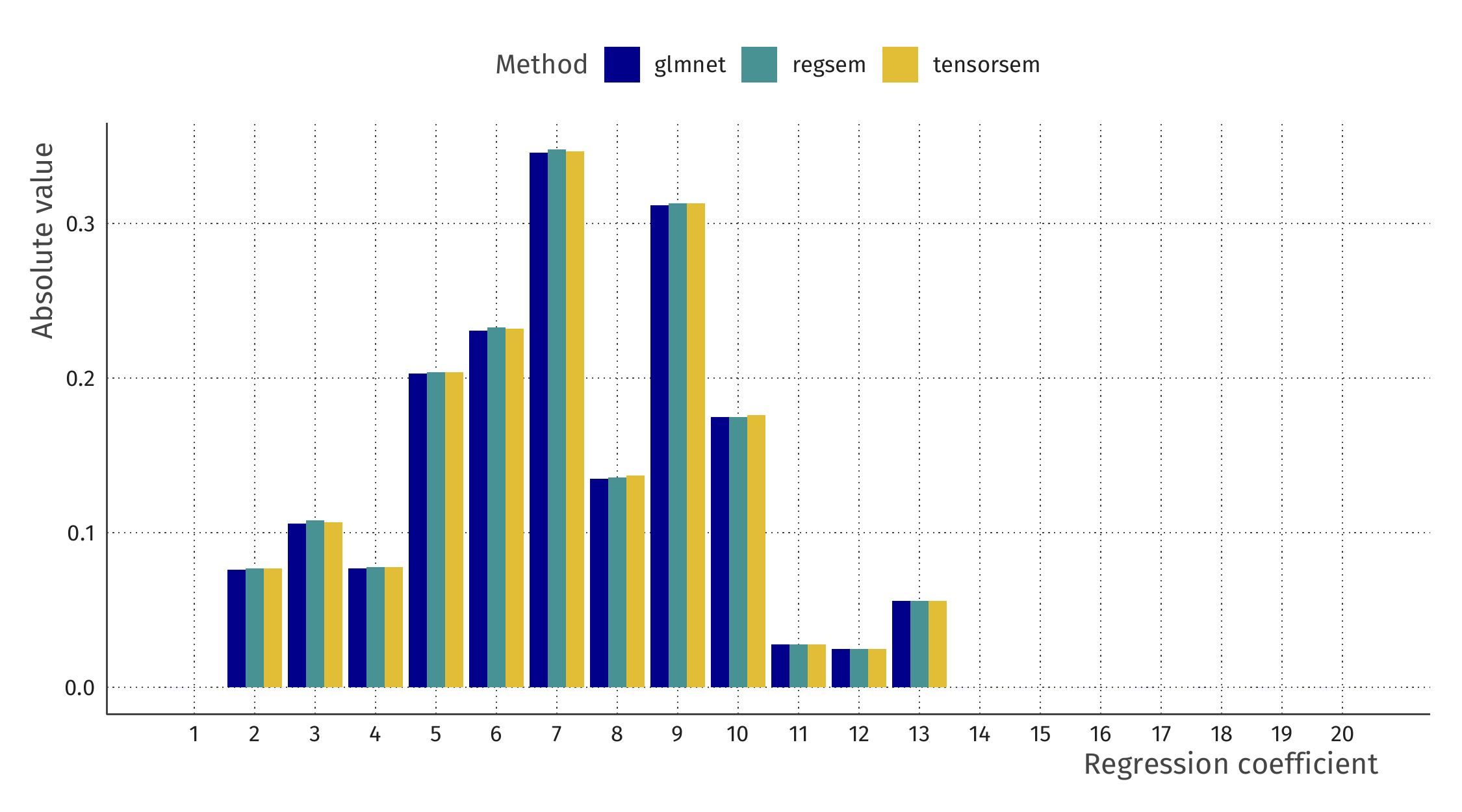} 

}

\caption{Absolute value of parameter estimates for a LASSO penalized regression model with 20 predictors. Here, \texttt{tensorsem} is compared to existing approaches and shown to provide similar parameter estimates.}\label{fig:res-lasso}
\end{figure}

In this example, we show how LASSO penalization on the regression parameters in \texttt{tensorsem} compares to \texttt{regsem} (Jacobucci et al., 2016) and \texttt{glmnet} (Friedman et al., 2010). For this, we generate data with a sample size of 1000 from a regression model with a single outcome variable, 10 true predictors, and 10 unrelated variables. The resulting parameter estimates for the three different estimation methods are shown in Figure \ref{fig:res-lasso}. The figure shows that with the chosen penalty parameter (0.11 for \texttt{regsem} and \texttt{tensorsem}, 0.028 for \texttt{glmnet} due to a difference in scaling), the estimates are very close in value. As expected, some parameters are shrunk to 0 for all three methods.

\begin{figure}[H]

{\centering \includegraphics[width=1\linewidth]{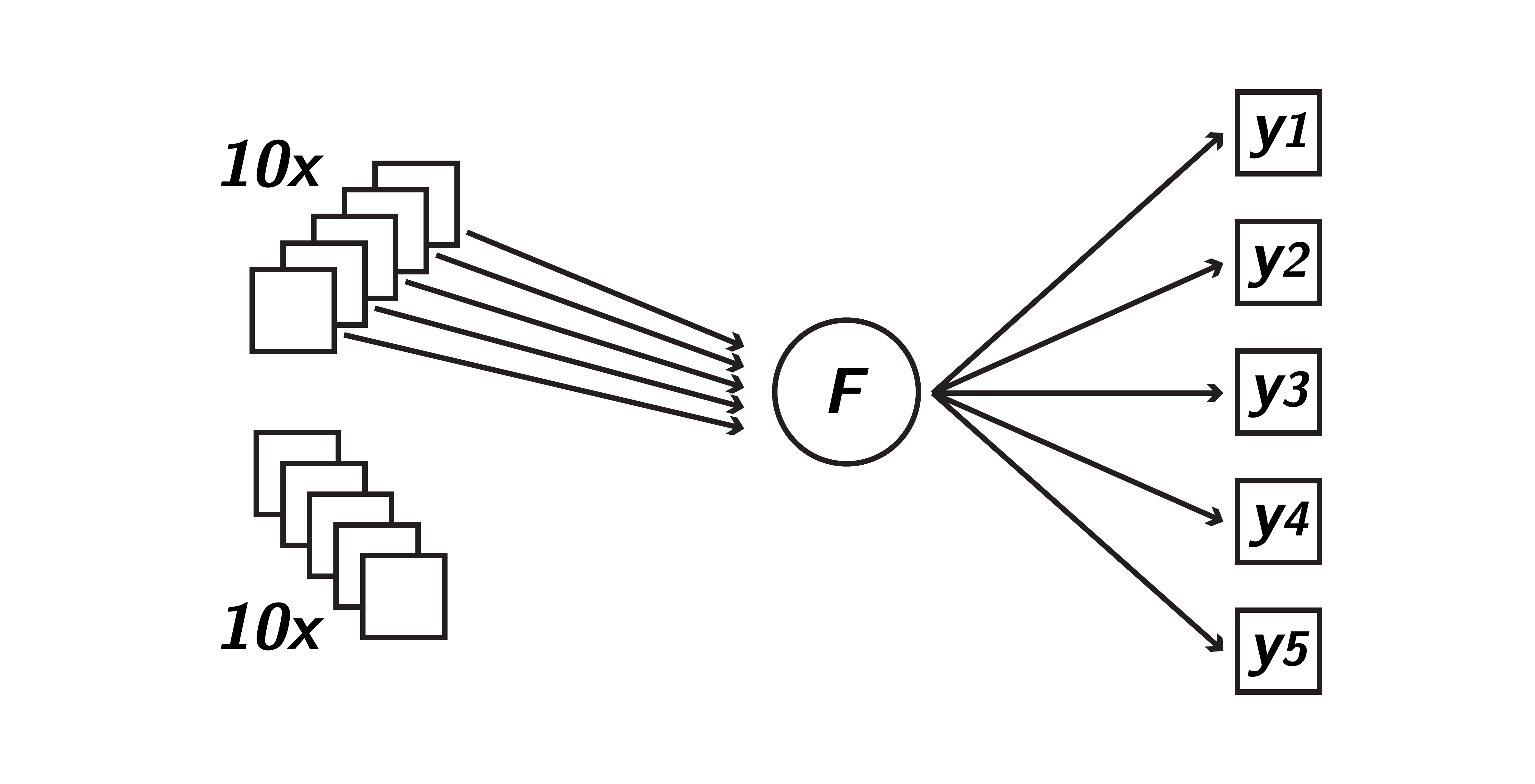} 

}

\caption{MIMIC model used to generate data to perform and compare LASSO estimation across \texttt{regsem}, and \texttt{tensorsem}. There are 10 predictors and 10 unrelated variables, and the latent outcome variable has 5 indicators. For clarity, residual variances of the indicators have been omitted from the figure.}\label{fig:ex-mimic}
\end{figure}

The advantage of regularized \(\beta\) parameters in SEM over the regularized GLM approach is that the outcome variable can also be latent as in a multiple-indicators multiple-causes (MIMIC) model. Jacobucci et al. (2018) demonstrate that such a model can be useful to relate high-dimensional features of brain images to cognitive measurements of a single or multiple underlying phenomena such as depression, anxiety, and stress. Following the example of Jacobucci et al. (2018), we simulated data from a MIMIC model with sparse causes and known indicators (Figure \ref{fig:ex-mimic}) and compared the solutions of the \texttt{tensorsem} approach to \texttt{regsem}.

\begin{figure}[H]

{\centering \includegraphics[width=1\linewidth]{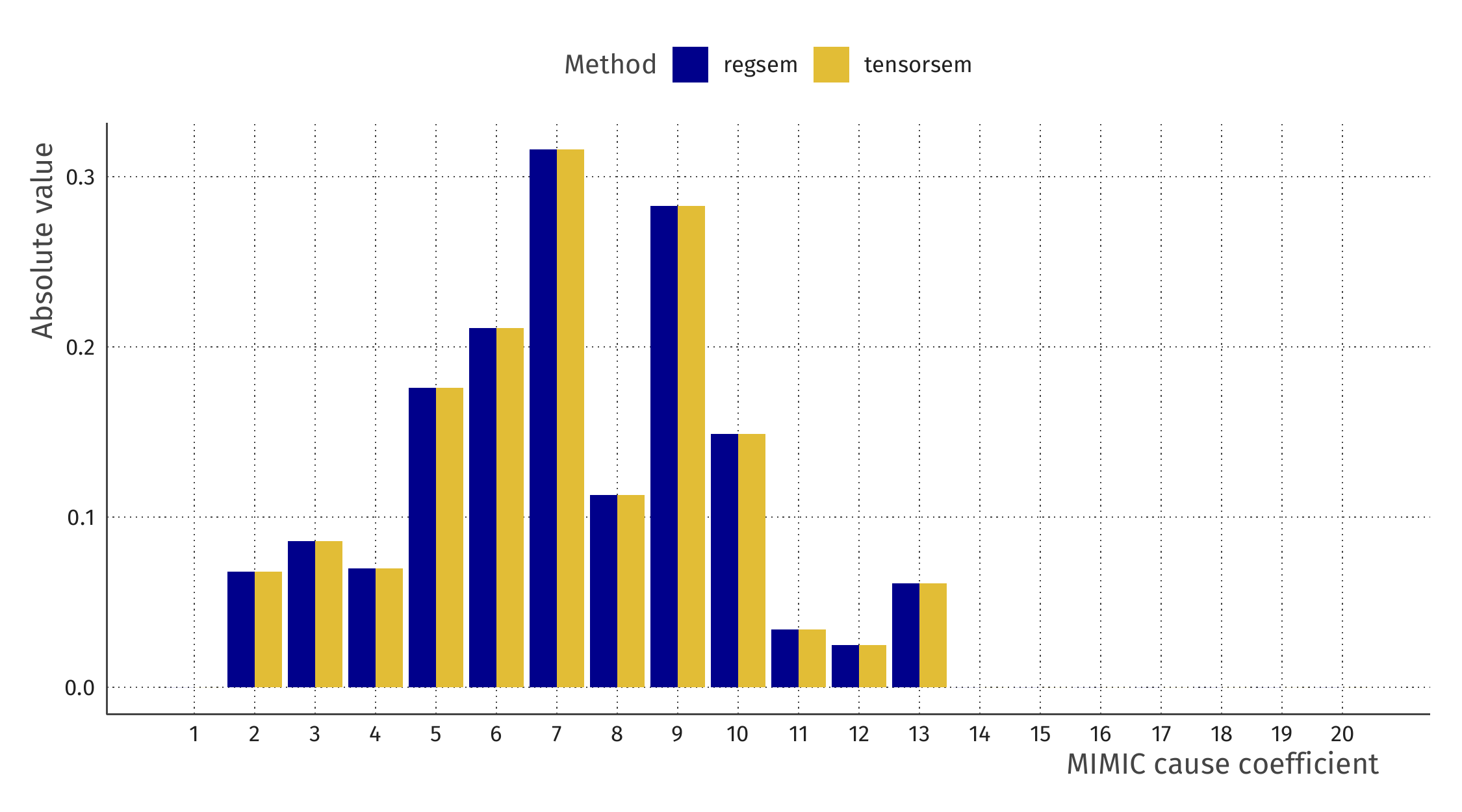} 

}

\caption{Absolute value of parameter estimates for a LASSO penalized MIMIC model with 20 predictors. The \texttt{tensorsem} implementation is compared to the existing approach \texttt{regsem} and shown to provide similar parameter estimates.}\label{fig:res-mimic}
\end{figure}

The results, displayed in Figure \ref{fig:res-mimic} show that the \(\beta\) coefficients are the same across \texttt{regsem} and \texttt{tensorsem}. This indicates that the suggestions from the tutorial paper by Jacobucci et al. (2018) for these models apply to our approach. In conclusion, this example shows that LASSO penalties can be imposed on the regression coefficients (\(\boldsymbol{B}_0\)) in SEM using the computation graph approach, that these parameters can be estimated using Adam, and that the resulting parameter estimates closely resemble those from existing approaches.

\hypertarget{sec:regfa}{%
\subsection{Sparse factor analysis}\label{sec:regfa}}

Obtaining sparsity in factor analysis is a large and old field of research, with methods including rotations of factor solutions in principal component analysis (Kaiser, 1958) and modification indices in CFA (Saris, Satorra, \& Sörbom, 1987; Sörbom, 1989). Sparsity is desirable in factor analysis due to the enhanced interpretability of the obtained factors. Recently, penalization has been applied to different factor analysis situations in order to obtain sparse factor loadings and simple solutions (Jin et al., 2018; Pan, Ip, \& Dubé, 2017; Scharf \& Nestler, 2019). Following these recent developments, in this last example we impose sparse structure in a CFA by imposing a penalty on the \(\boldsymbol{\Lambda}\) matrix. Here, we use both the LASSO and the spike-and-slab penalty from equation \eqref{eq:spikeslab}.

\begin{figure}[H]

{\centering \includegraphics[width=1\linewidth]{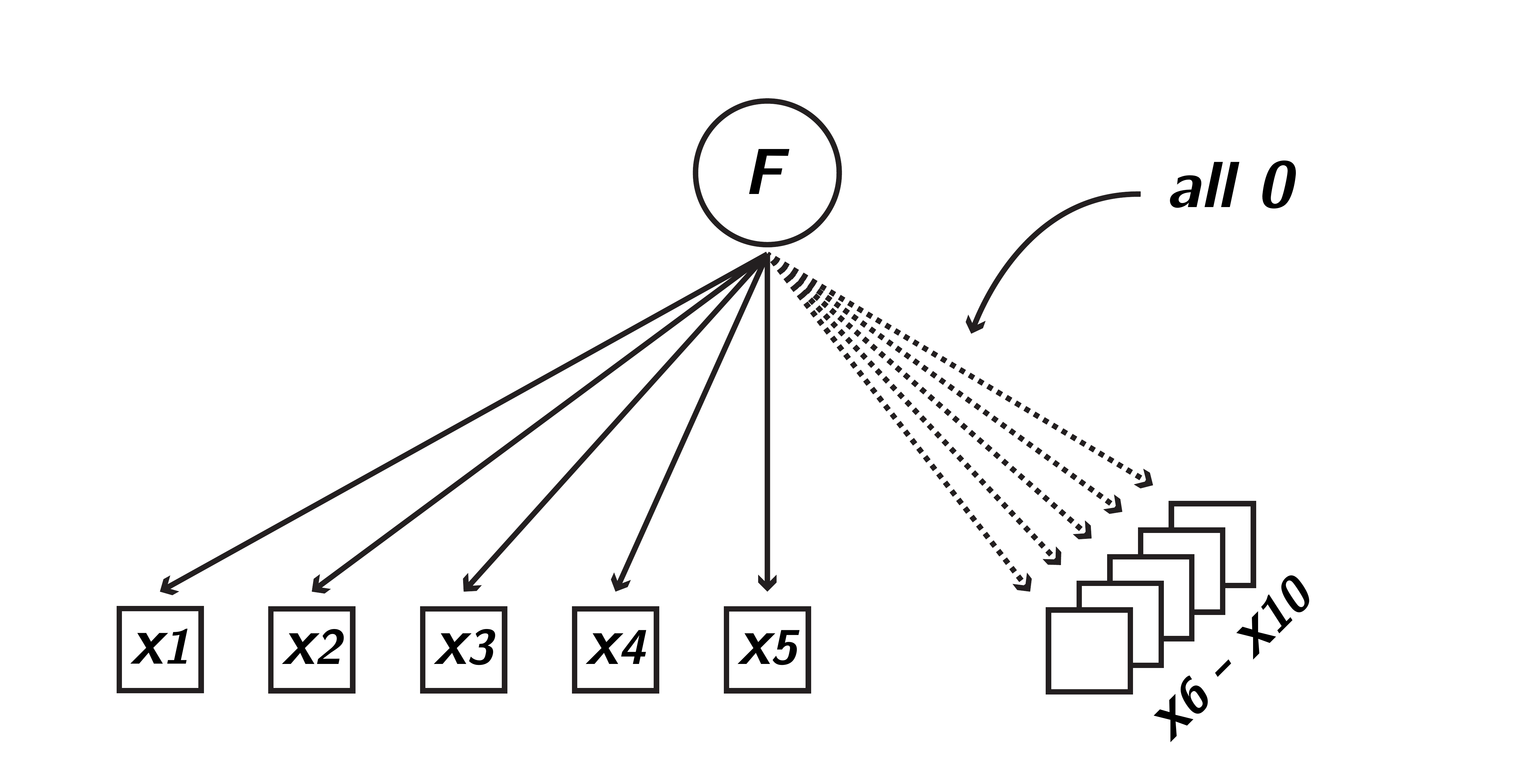} 

}

\caption{Factor analysis population model used generate data for the penalized \(\boldsymbol{\Lambda}\) matrix sparse factor analysis procedures. There are 5 true indicators with nonzero factor loadings, and 5 unrelated variables.}\label{fig:ex-efa}
\end{figure}

In this example, we show the effect of ML estimation with \(\boldsymbol{\Lambda}\) penalization on a dataset with 300 samples generated from the mechanism shown in Figure \ref{fig:ex-efa}. The results of the penalized factor analysis are displayed in Table \ref{tab:res-efa}. As the penalty parameter \(\lambda\) is increased from 0, the factor loadings are shrunk more, with some being set to 0. With a LASSO penalty of strength 0.3, the true data-generating structure is recovered. Note that although the difference is small in this low-dimensional situation, the spike-and-slab prior leads to the same structure with slightly less bias when compared to the ML solution (\(\lambda = 0\)). This is in line with the findings of van Erp et al. (2019). For this result, we have set the hyperparameters of the \(F_{\text{SS}}\) objective from Equation \eqref{eq:spikeslab} to \(\lambda_1 = 0.55, \lambda_2 = 0.05, \pi = 0.5\).

\begin{table}[H]

\caption{\label{tab:res-efa}Results for penalized factor analysis. Sparsity is induced as the penalty parameter \(\lambda\) is raised from 0 (ML estimates) to 0.1 to 0.3, where the data-generating structure is recovered. The spike-and-slab penalty leads to the same structure with slightly less bias in the nonzero factor loadings.}
\centering
\begin{tabular}[t]{llccccccccc}
\toprule
  & X1 & X2 & X3 & X4 & X5 & X6 & X7 & X8 & X9 & X10\\
\midrule
$\lambda$ = 0 & 1.075 & 0.728 & -0.323 & -0.482 & -0.811 & 0.116 & 0.015 & 0.095 & -0.003 & 0.007\\
$\lambda$ = 0.1 & 0.942 & 0.649 & -0.294 & -0.441 & -0.720 & 0.056 & 0.000 & 0.036 & 0.000 & 0.000\\
$\lambda$ = 0.3 & 0.754 & 0.545 & -0.258 & -0.393 & -0.596 & 0.000 & 0.000 & 0.000 & 0.000 & 0.000\\
spike-slab & 0.761 & 0.553 & -0.262 & -0.398 & -0.604 & 0.000 & 0.000 & 0.000 & 0.000 & 0.000\\
\bottomrule
\end{tabular}
\end{table}

This example shows that sparsity in the factor loadings can be imposed in a factor analysis with a penalty on the \(\boldsymbol{\Lambda}\) matrix. If the assumption of simple structure holds, the penalized computation graph approach may recover this structure with an appropriate amount of penalization.

In the \texttt{tensorsem} package, the strength of the penalties can be changed during training. For example, a researcher may arrive at the ML solution and from there slowly increase the penalty strength and inspect different solutions. In this way, the entire model need not be estimated for each value of the \(\lambda\) parameter. This iterative procedure is similar to the path procedure of Friedman et al. (2010) which makes \texttt{glmnet} an efficient method for penalization.

Together, the examples have shown that estimation and extension of SEM through computation graphs is viable.

\hypertarget{conclusion}{%
\section{Conclusion}\label{conclusion}}

Estimation of SEM becomes more challenging as latent variable models become larger and more complex. Traditionally, SEM optimizers have already suffered from nonconvergence and inadmissible solutions (e.g., Chen et al., 2001; Revilla \& Saris, 2013), and with the increasing complexity of available datasets these problems are set to become more relevant. We argue that current estimation methods do not fulfil the needs of researchers applying SEM to novel situations in the future.

In this paper, we have introduced a new way of constructing objective functions for SEM by using computation graphs. When combined with a modern optimizer such as Adam, available in the software package \texttt{TensorFlow}, this approach opens up new directions for SEM estimation. The flexibility of the computation graph lies in the ease with which the graph is edited, after which gradients are computed automatically and optimization can be performed without in-depth mathematical analysis. This holds even for non-convex objectives and objectives which are not continuously differentiable. With this tool, SEM researchers can quickly implement extensions to SEM, such as automatic variable selection in MIMIC models (Jacobucci et al., 2018) or sparse factor models (Jin et al., 2018).

We have shown that various previously proposed improvements to SEM follow naturally from this framework, and that our implementation called \texttt{tensorsem} is able to optimize these, yielding parameter estimates that behave according to expectations. In addition, we demonstrated the ease with which extensions can be investigated by introducing the spike-and-slab penalty, which is a novel penalization method for SEM. Going forward, \texttt{tensorsem} will be developed as a flexible alternative SEM software package aimed at researchers who want to apply SEM to novel situations.

As the computation graph approach paves the way for a more flexible SEM, researchers can use it to develop theoretical SEM improvements. For example, future research can focus on how penalties may be used to improve the performance and interpretability of specific models (e.g., Jacobucci et al., 2018), or how different objective functions may be used to bring SEM to novel situations such as high-dimensional data (Grotzinger et al., 2018; van Kesteren, 2019). A potential extension to SEM is the use of high-dimensional covariates to debias inferences in observational studies (Athey, Imbens, \& Wager, 2018). The computation graph may aid in importing such procedures to SEM. An interesting historical note is that Cudeck et al. (1993) have had similar reasons for creating a general SEM optimization program, where the full Hessian is numerically approximated for any covariance model and the solution is computed using Gauss-Newton iterations. Modern computational tools now make such generic SEM programs feasible.

Another topic for future research is exploratory model specification. For example, Brandmaier, Oertzen, McArdle, \& Lindenberger (2013) and Brandmaier, Prindle, McArdle, \& Lindenberger (2016) use decision trees to find relevant covariates in SEM, and Marcoulides \& Drezner (2001) use genetic algorithms to perform model specification search. Penalties provide a natural way to automatically set some parameters to 0, which is equivalent to specifying constraints in the model. A compelling example of this is the work by Pan et al. (2017), who used the Bayesian form of LASSO regularization as an alternative to post-hoc model modification in CFA. Their approach penalizes the residual covariance matrix of the indicators, leading to a more sparse selection of residual covariance parameters to be freed relative to the common modification index approach.

There is a need for the SEM computation graph approach to be further developed in terms of speed and capabilities in order to expand its range of applications. For example, through applying Adam as a stochastic gradient descent (SGD) optimizer it may be extended to perform full information maximum likelihood (FIML) estimation, batch-wise estimation, or SEM estimation with millions of observations. This will potentially enable SEM to be performed on completely novel types of data, such as streaming data, images, or sounds. Another improvement which may be imported from the deep learning literature is computation of approximate Bayesian posterior credible intervals for any objective function using stochastic gradient descent steps at the optimum (Mandt, Hoffman, \& Blei, 2017). The deep learning optimization literature moves fast, and through the connections we have established in this paper the SEM literature could benefit from its pace.

\hypertarget{acknowledgments}{%
\section*{Acknowledgments}\label{acknowledgments}}
\addcontentsline{toc}{section}{Acknowledgments}

This work was supported by the Netherlands Organization for Scientific Research (NWO) under grant number 406.17.057.

We thank Rogier Kievit for his comments on an earlier version of this manuscript, and Maksim Rudnev for his helpful questions regarding \texttt{tensorsem}.

\hypertarget{references}{%
\section*{References}\label{references}}
\addcontentsline{toc}{section}{References}

\begingroup
\setlength{\parindent}{-0.5in}
\setlength{\leftskip}{0.5in}

\noindent

\hypertarget{refs}{}
\leavevmode\hypertarget{ref-Abadi2016}{}%
Abadi, M., Barham, P., Chen, J., Chen, Z., Davis, A., Dean, J., \ldots{} others. (2016). Tensorflow: A system for large-scale machine learning. In \emph{12th \(\{\)usenix\(\}\) symposium on operating systems design and implementation (\(\{\)osdi\(\}\) 16)} (pp. 265--283).

\leavevmode\hypertarget{ref-Allaire2019}{}%
Allaire, J., \& Tang, Y. (2019). \emph{tensorflow: R interface to 'TensorFlow'}. Retrieved from \url{https://CRAN.R-project.org/package=tensorflow}

\leavevmode\hypertarget{ref-Asparouhov2018}{}%
Asparouhov, T., Hamaker, E. L., \& Muthén, B. (2018). Dynamic structural equation models. \emph{Structural Equation Modeling: A Multidisciplinary Journal}, \emph{25}(3), 359--388.

\leavevmode\hypertarget{ref-Athey2018}{}%
Athey, S., Imbens, G. W., \& Wager, S. (2018). Approximate residual balancing: Debiased inference of average treatment effects in high dimensions. \emph{Journal of the Royal Statistical Society: Series B (Statistical Methodology)}, \emph{80}(4), 597--623.

\leavevmode\hypertarget{ref-Bengio2000}{}%
Bengio, Y. (2000). Gradient-based optimization of hyperparameters. \emph{Neural Computation}, \emph{12}(8), 1889--1900.

\leavevmode\hypertarget{ref-Bentler1980}{}%
Bentler, P. M., \& Weeks, D. G. (1980). Linear structural equations with latent variables. \emph{Psychometrika}, \emph{45}(3), 289--308.

\leavevmode\hypertarget{ref-Betancourt2017}{}%
Betancourt, M. (2017). A conceptual introduction to hamiltonian monte carlo. \emph{arXiv Preprint arXiv:1701.02434}.

\leavevmode\hypertarget{ref-Bock1966}{}%
Bock, R. D., \& Bargmann, R. E. (1966). Analysis of covariance structures. \emph{Psychometrika}, \emph{31}(4), 507--534.

\leavevmode\hypertarget{ref-Bollen1989}{}%
Bollen, K. A. (1989). Structural equations with latent variables.

\leavevmode\hypertarget{ref-Brandmaier2013}{}%
Brandmaier, A. M., Oertzen, T. von, McArdle, J. J., \& Lindenberger, U. (2013). Structural equation model trees. \emph{Psychological Methods}, \emph{18}(1), 71.

\leavevmode\hypertarget{ref-Brandmaier2016}{}%
Brandmaier, A. M., Prindle, J. J., McArdle, J. J., \& Lindenberger, U. (2016). Theory-guided exploration with structural equation model forests. \emph{Psychological Methods}, \emph{21}(4), 566--582. \url{https://doi.org/10.1037/met0000090}

\leavevmode\hypertarget{ref-Browne1974}{}%
Browne, M. W. (1974). Generalized least squares estimators in the analysis of covariance structures. \emph{South African Statistical Journal}, \emph{8}(1), 1--24.

\leavevmode\hypertarget{ref-Byrne2016}{}%
Byrne, B. M. (2016). \emph{Structural equation modeling with amos: Basic concepts, applications, and programming}. Routledge.

\leavevmode\hypertarget{ref-Carpenter2017}{}%
Carpenter, B., Gelman, A., Hoffman, M. D., Lee, D., Goodrich, B., Betancourt, M., \ldots{} Riddell, A. (2017). Stan: A probabilistic programming language. \emph{Journal of Statistical Software}, \emph{76}(1).

\leavevmode\hypertarget{ref-Cernat2019}{}%
Cernat, A., \& Oberski, D. L. (2019). Extending the within-persons experimental design: The multitrait-multierror (MTME) approach. In P. J. Lavrakas, M. W. Traugott, C. Kennedy, A. L. Holbrook, \& E. de Leeuw (Eds.), \emph{Experimental methods in survey research: Techniques that combine random sampling with random assignment}. New York: John Wiley \& Sons.

\leavevmode\hypertarget{ref-Chen2001}{}%
Chen, F., Bollen, K. A., Paxton, P., Curran, P. J., \& Kirby, J. B. (2001). Improper solutions in structural equation models: Causes, consequences, and strategies. \emph{Sociological Methods \& Research}, \emph{29}(4), 468--508.

\leavevmode\hypertarget{ref-Collobert2002}{}%
Collobert, R., Bengio, S., \& Mariéthoz, J. (2002). \emph{Torch: A modular machine learning software library}. Idiap.

\leavevmode\hypertarget{ref-cudeck1993simple}{}%
Cudeck, R., Klebe, K. J., \& Henly, S. J. (1993). A simple gauss-newton procedure for covariance structure analysis with high-level computer languages. \emph{Psychometrika}, \emph{58}(2), 211--232.

\leavevmode\hypertarget{ref-Epskamp2018}{}%
Epskamp, S., Borsboom, D., \& Fried, E. I. (2018). Estimating psychological networks and their accuracy: A tutorial paper. \emph{Behavior Research Methods}, \emph{50}(1), 195--212.

\leavevmode\hypertarget{ref-epskamp2017generalized}{}%
Epskamp, S., Rhemtulla, M., \& Borsboom, D. (2017). Generalized network psychometrics: Combining network and latent variable models. \emph{Psychometrika}, \emph{82}(4), 904--927.

\leavevmode\hypertarget{ref-Fan2001}{}%
Fan, J., \& Li, R. (2001). Variable selection via nonconcave penalized likelihood and its oracle properties. \emph{Journal of the American Statistical Association}, \emph{96}(456), 1348--1360.

\leavevmode\hypertarget{ref-Friedman2010}{}%
Friedman, J., Hastie, T., \& Tibshirani, R. (2010). Regularization paths for generalized linear models via coordinate descent. \emph{Journal of Tatistical Software}, \emph{33}(1), 1.

\leavevmode\hypertarget{ref-gabry2019visualization}{}%
Gabry, J., Simpson, D., Vehtari, A., Betancourt, M., \& Gelman, A. (2019). Visualization in bayesian workflow. \emph{Journal of the Royal Statistical Society: Series A (Statistics in Society)}, \emph{182}(2), 389--402.

\leavevmode\hypertarget{ref-goeman2018l1}{}%
Goeman, J., Meijer, R., \& Chaturvedi, N. (2018). L1 and l2 penalized regression models. \emph{Vignette R Package Penalized.}

\leavevmode\hypertarget{ref-Goodfellow2016}{}%
Goodfellow, I., Bengio, Y., \& Courville, A. (2016). \emph{Deep learning}. MIT Press.

\leavevmode\hypertarget{ref-grotzinger2019genomic}{}%
Grotzinger, A. D., Rhemtulla, M., Vlaming, R. de, Ritchie, S. J., Mallard, T. T., Hill, W. D., \ldots{} others. (2019). Genomic structural equation modelling provides insights into the multivariate genetic architecture of complex traits. \emph{Nature Human Behaviour}, \emph{3}(5), 513.

\leavevmode\hypertarget{ref-Grotzinger2018}{}%
Grotzinger, A. D., Rhemtulla, M., Vlaming, R. de, Ritchie, S. J., Mallard, T. T., Hill, W. D., \ldots{} others. (2018). Genomic sem provides insights into the multivariate genetic architecture of complex traits. \emph{BioRxiv}, 305029.

\leavevmode\hypertarget{ref-Hastie2015}{}%
Hastie, T., Tibshirani, R., \& Wainwright, M. (2015). \emph{Statistical Learning with Sparsity: The Lasso and Generalizations} (p. 362). Boca Raton: CRC Press. \url{https://doi.org/10.1201/b18401-1}

\leavevmode\hypertarget{ref-Helm2017}{}%
Helm, J. L., Castro-Schilo, L., \& Oravecz, Z. (2017). Bayesian versus maximum likelihood estimation of multitrait--multimethod confirmatory factor models. \emph{Structural Equation Modeling: A Multidisciplinary Journal}, \emph{24}(1), 17--30.

\leavevmode\hypertarget{ref-Holzinger1939}{}%
Holzinger, K. J., \& Swineford, F. (1939). A study in factor analysis: The stability of a bi-factor solution. \emph{Supplementary Educational Monographs}.

\leavevmode\hypertarget{ref-Jacobucci2017}{}%
Jacobucci, R. (2017). regsem: Regularized Structural Equation Modeling. \emph{ArXiv Preprint}. Retrieved from \url{http://arxiv.org/abs/1703.08489}

\leavevmode\hypertarget{ref-Jacobucci2018a}{}%
Jacobucci, R., Brandmaier, A. M., \& Kievit, R. A. (2018). Variable selection in structural equation models with regularized MIMIC models. \emph{PsyArXiv Preprint}, 1--40. \url{https://doi.org/10.17605/OSF.IO/BXZJF}

\leavevmode\hypertarget{ref-Jacobucci2016}{}%
Jacobucci, R., Grimm, K. J., \& McArdle, J. J. (2016). Regularized structural equation modeling. \emph{Structural Equation Modeling}, \emph{23}(4), 555--566. \url{https://doi.org/10.1080/10705511.2016.1154793.Regularized}

\leavevmode\hypertarget{ref-Jin2018}{}%
Jin, S., Moustaki, I., \& Yang-Wallentin, F. (2018). Approximated penalized maximum likelihood for exploratory factor analysis: An orthogonal case. \emph{Psychometrika}, \emph{83}(3), 628--649.

\leavevmode\hypertarget{ref-Joreskog1966}{}%
Jöreskog, K. G. (1966). Testing a simple structure hypothesis in factor analysis. \emph{Psychometrika}, \emph{31}(2), 165--178.

\leavevmode\hypertarget{ref-Joreskog1967}{}%
Jöreskog, K. G. (1967). Some contributions to maximum likelihood factor analysis. \emph{Psychometrika}, \emph{32}(4), 443--482.

\leavevmode\hypertarget{ref-Joreskog1993}{}%
Jöreskog, K. G., \& Sörbom, D. (1993). \emph{LISREL 8: Structural equation modeling with the simplis command language}. Scientific Software International.

\leavevmode\hypertarget{ref-Kaiser1958}{}%
Kaiser, H. F. (1958). The varimax criterion for analytic rotation in factor analysis. \emph{Psychometrika}, \emph{23}(3), 187--200.

\leavevmode\hypertarget{ref-Kingma2014}{}%
Kingma, D. P., \& Ba, J. (2014). Adam: A method for stochastic optimization. \emph{arXiv Preprint arXiv:1412.6980}.

\leavevmode\hypertarget{ref-Kolenikov2012}{}%
Kolenikov, S., \& Bollen, K. A. (2012). Testing negative error variances: Is a heywood case a symptom of misspecification? \emph{Sociological Methods \& Research}, \emph{41}(1), 124--167.

\leavevmode\hypertarget{ref-kyung2010penalized}{}%
Kyung, M., Gill, J., Ghosh, M., Casella, G., \& others. (2010). Penalized regression, standard errors, and bayesian lassos. \emph{Bayesian Analysis}, \emph{5}(2), 369--411.

\leavevmode\hypertarget{ref-Lee1979}{}%
Lee, S.-Y., \& Jennrich, R. (1979). A study of algorithms for covariance structure analysis with specific comparisons using factor analysis. \emph{Psychometrika}, \emph{44}(1), 99--113.

\leavevmode\hypertarget{ref-Mandt2017}{}%
Mandt, S., Hoffman, M. D., \& Blei, D. M. (2017). Stochastic gradient descent as approximate bayesian inference. \emph{The Journal of Machine Learning Research}, \emph{18}(1), 4873--4907.

\leavevmode\hypertarget{ref-Marcoulides2001}{}%
Marcoulides, G. A., \& Drezner, Z. (2001). Specification searches in structural equation modeling with a genetic algorithm. \emph{New Developments and Techniques in Structural Equation Modeling}, 247--268.

\leavevmode\hypertarget{ref-McArdle1984}{}%
McArdle, J. J., \& McDonald, R. P. (1984). Some algebraic properties of the reticular action model for moment structures. \emph{British Journal of Mathematical and Statistical Psychology}, \emph{37}(2), 234--251.

\leavevmode\hypertarget{ref-Merkle2015}{}%
Merkle, E. C., \& Rosseel, Y. (2015). Blavaan: Bayesian structural equation models via parameter expansion. \emph{arXiv Preprint arXiv:1511.05604}.

\leavevmode\hypertarget{ref-Muthen2012}{}%
Muthén, B., \& Asparouhov, T. (2012). Bayesian Structural Equation Modeling: A More Flexible Representation of Substantive Theory. \emph{Psychological Methods}, \emph{17}(3), 313--335. \url{https://doi.org/10.1037/a0026802}

\leavevmode\hypertarget{ref-Muthen2019}{}%
Muthén, L., \& Muthén, B. (2019). Mplus. \emph{The Comprehensive Modelling Program for Applied Researchers: User's Guide}, \emph{5}.

\leavevmode\hypertarget{ref-Neale2016}{}%
Neale, M. C., Hunter, M. D., Pritikin, J. N., Zahery, M., Brick, T. R., Kirkpatrick, R. M., \ldots{} Boker, S. M. (2016). OpenMx 2.0: Extended structural equation and statistical modeling. \emph{Psychometrika}, \emph{81}(2), 535--549. \url{https://doi.org/10.1007/s11336-014-9435-8}

\leavevmode\hypertarget{ref-Neudecker1991}{}%
Neudecker, H., \& Satorra, A. (1991). Linear structural relations: Gradient and hessian of the fitting function. \emph{Statistics \& Probability Letters}, \emph{11}(1), 57--61.

\leavevmode\hypertarget{ref-VonOertzen2015}{}%
Oertzen, T. von, Brandmaier, A. M., \& Tsang, S. (2015). Structural equation modeling with \(\Omega\)nyx. \emph{Structural Equation Modeling: A Multidisciplinary Journal}, \emph{22}(1), 148--161.

\leavevmode\hypertarget{ref-Pan2017}{}%
Pan, J., Ip, E. H., \& Dubé, L. (2017). An alternative to post hoc model modification in confirmatory factor analysis: The bayesian lasso. \emph{Psychological Methods}, \emph{22}(4), 687.

\leavevmode\hypertarget{ref-Paszke2017}{}%
Paszke, A., Gross, S., Chintala, S., Chanan, G., Yang, E., DeVito, Z., \ldots{} Lerer, A. (2017). Automatic differentiation in pytorch.

\leavevmode\hypertarget{ref-RCoreTeam2018}{}%
R Core Team. (2018). R: A language and environment for statistical computing. Vienna, Austria: R Foundation for Statistical Computing. Retrieved from \url{https://www.r-project.org/}

\leavevmode\hypertarget{ref-Revilla2013}{}%
Revilla, M., \& Saris, W. E. (2013). The split-ballot multitrait-multimethod approach: Implementation and problems. \emph{Structural Equation Modeling: A Multidisciplinary Journal}, \emph{20}(1), 27--46.

\leavevmode\hypertarget{ref-Rindskopf1983}{}%
Rindskopf, D. (1983). Parameterizing inequality constraints on unique variances in linear structural models. \emph{Psychometrika}, \emph{48}(1), 73--83.

\leavevmode\hypertarget{ref-Rindskopf1984}{}%
Rindskopf, D. (1984). Structural equation models: Empirical identification, heywood cases, and related problems. \emph{Sociological Methods \& Research}, \emph{13}(1), 109--119.

\leavevmode\hypertarget{ref-Rockova2018}{}%
Ročková, V., \& George, E. I. (2018). The spike-and-slab lasso. \emph{Journal of the American Statistical Association}, \emph{113}(521), 431--444.

\leavevmode\hypertarget{ref-Rosseel2012}{}%
Rosseel, Y. (2012). Lavaan: An R package for structural equation modeling. \emph{Journal of Statistical Software}, \emph{48}(2), 1--36.

\leavevmode\hypertarget{ref-Saris1987}{}%
Saris, W. E., Satorra, A., \& Sörbom, D. (1987). The detection and correction of specification errors in structural equation models. \emph{Sociological Methodology}, 105--129.

\leavevmode\hypertarget{ref-Satorra1988}{}%
Satorra, A., \& Bentler, P. (1988). Scaling corrections for statistics in covariance structure analysis.

\leavevmode\hypertarget{ref-savalei2014understanding}{}%
Savalei, V. (2014). Understanding robust corrections in structural equation modeling. \emph{Structural Equation Modeling: A Multidisciplinary Journal}, \emph{21}(1), 149--160.

\leavevmode\hypertarget{ref-Scardapane2017}{}%
Scardapane, S., Comminiello, D., Hussain, A., \& Uncini, A. (2017). Group sparse regularization for deep neural networks. \emph{Neurocomputing}, \emph{241}, 81--89.

\leavevmode\hypertarget{ref-scharf2019should}{}%
Scharf, F., \& Nestler, S. (2019). Should regularization replace simple structure rotation in exploratory factor analysis? \emph{Structural Equation Modeling: A Multidisciplinary Journal}, 1--15.

\leavevmode\hypertarget{ref-sclove1987application}{}%
Sclove, S. L. (1987). Application of model-selection criteria to some problems in multivariate analysis. \emph{Psychometrika}, \emph{52}(3), 333--343.

\leavevmode\hypertarget{ref-Siemsen2007}{}%
Siemsen, E., \& Bollen, K. A. (2007). Least absolute deviation estimation in structural equation modeling. \emph{Sociological Methods \& Research}, \emph{36}(2), 227--265.

\leavevmode\hypertarget{ref-Sorbom1989}{}%
Sörbom, D. (1989). Model modification. \emph{Psychometrika}, \emph{54}(3), 371--384.

\leavevmode\hypertarget{ref-Stata2017}{}%
StataCorp. (2017). Stata statistical software: Release 15.

\leavevmode\hypertarget{ref-Tieleman2012}{}%
Tieleman, T., \& Hinton, G. (2012). Lecture 6.5-rmsprop: Divide the gradient by a running average of its recent magnitude. \emph{COURSERA: Neural Networks for Machine Learning}, \emph{4}(2), 26--31.

\leavevmode\hypertarget{ref-VanDeGeer2009}{}%
Van De Geer, S. A., Bühlmann, P., \& others. (2009). On the conditions used to prove oracle results for the lasso. \emph{Electronic Journal of Statistics}, \emph{3}, 1360--1392.

\leavevmode\hypertarget{ref-Erp2019}{}%
van Erp, S., Oberski, D. L., \& Mulder, J. (2019). Shrinkage Priors for Bayesian Penalized Regression. \emph{Journal of Mathematical Psychology}, \emph{89}, 31--50. \url{https://doi.org/10.31219/osf.io/cg8fq}

\leavevmode\hypertarget{ref-vanKesteren2019a}{}%
van Kesteren, E.-J. (2019). \emph{tensorsem: Structural equation models using TensorFlow}.

\leavevmode\hypertarget{ref-vanKesteren2019b}{}%
van Kesteren, E.-J., \& Oberski, D. L. (2019). Exploratory mediation analysis with many potential mediators. \emph{Structural Equation Modeling: A Multidisciplinary Journal}, 1--14.

\leavevmode\hypertarget{ref-vehtari2017practical}{}%
Vehtari, A., Gelman, A., \& Gabry, J. (2017). Practical bayesian model evaluation using leave-one-out cross-validation and waic. \emph{Statistics and Computing}, \emph{27}(5), 1413--1432.

\leavevmode\hypertarget{ref-Voelkle2013}{}%
Voelkle, M. C., \& Oud, J. H. (2013). Continuous time modelling with individually varying time intervals for oscillating and non-oscillating processes. \emph{British Journal of Mathematical and Statistical Psychology}, \emph{66}(1), 103--126.

\leavevmode\hypertarget{ref-Wengert1964}{}%
Wengert, R. E. (1964). A simple automatic derivative evaluation program. \emph{Communications of the ACM}, \emph{7}(8), 463--464.

\leavevmode\hypertarget{ref-Zhang2010}{}%
Zhang, C.-H. (2010). Nearly unbiased variable selection under minimax concave penalty. \emph{The Annals of Statistics}, \emph{38}(2), 894--942.

\leavevmode\hypertarget{ref-Zou2005}{}%
Zou, H., \& Hastie, T. (2005). Regularization and variable selection via the elastic-net. \emph{Journal of the Royal Statistical Society}, \emph{67}(2), 301--320. \url{https://doi.org/10.1111/j.1467-9868.2005.00503.x}

\endgroup

\end{document}